\documentstyle[12pt,epsf,epic,eepicemu]{article}
\def\inseps#1#2{\def\epsfsize##1##2{#2##1} \centerline{\epsfbox{#1}}}

\def\la{\langle}
\def\ra{\rangle}

\begin{document}

\title{Monte Carlo Results for Projected Self-Avoiding Polygons: 
A Two-dimensional Model for Knotted Polymers}

\author{E. Guitter and E. Orlandini \\
\vspace{0.025in}\\
 CEA-Saclay,
   Service de Physique Th\'eorique,\\ 
    F-91191 Gif-sur-Yvette Cedex, France \\
}

\maketitle

\vspace{5mm}

\unitlength = 8mm   
\newsavebox{\dashes}
\savebox{\dashes}(0.5,0){   
 \begin{picture}(0.5,0)     
   \thicklines
   \drawline(0,0)(0.13,0)
   \drawline(0.185,0)(0.315,0)
   \drawline(0.37,0)(0.5,0)
 \end{picture}
}
\def\Prob{{\rm Prob}}

\begin{abstract}
We introduce a two-dimensional lattice model for the description 
of knotted polymer rings. A polymer configuration is modeled by a closed 
polygon drawn on the square diagonal lattice, with possible crossings 
describing pairs of strands of polymer passing on top of each other. 
Each polygon configuration can be viewed as the two-dimensional projection
of a particular knot. We study numerically the statistics of large 
polygons with a fixed knot type, using a generalization of the BFACF 
algorithm for self-avoiding walks. This new algorithm incorporates
both the displacement of crossings and the three types of Reidemeister 
transformations preserving the knot topology. Its ergodicity within 
a fixed knot type is not proven here rigorously but strong arguments 
in favor of this ergodicity are given together with a tentative sketch
of proof. Assuming this ergodicity, we obtain numerically the following
results for the statistics of knotted polygons: In the limit of a low 
crossing fugacity, we find a localization along the polygon of all the 
primary factors forming the knot. Increasing the crossing fugacity gives 
rise to a transition from a self-avoiding walk to a branched polymer 
behavior.
\end{abstract}
\vfill\eject
The statistical properties of self-avoiding polymers are now very well
understood, mainly thanks to the famous equivalence of the problem 
to a field theory, leading to a natural derivation of scaling laws and 
critical exponents \cite{deGennes72}. In three dimensions however, 
{\it closed} polymers will
generally form knots, which, for self-avoiding objects, will restrict
the accessible configurations. In the last few years, some progress has been
made towards answering some basic questions about knots, showing for instance
that a sufficiently long polymer is knotted with probability one
\cite{SumWhi88,Pippenger89}, or computing the distribution of
random knots \cite{RCV93}. However, the statistics of a closed polymer
with a {\sl fixed} knot-type is less well understood. Indeed, the usual field 
theory approach does not account for this distinction of the knot type, 
but corresponds rather to a summation over all the possible knot topologies.
In particular, the natural question of how the critical exponents
depend on the knot topology remains an open issue. Knotted configurations
 are encountered in the description of closed DNA molecules, with an 
apparent influence of the knot type on some of their 
properties \cite{SW93}.

An attempt to understand the role of a fixed knot topology was done 
in \cite{JW91a,DT93,DT94,Quake94,OTJW96,OTJW98} where a numerical study of 
knotted lattice 
polygons in the cubic lattice $Z^3$ was performed. There it was shown that, while 
the connectivity constant and the size exponent $\nu$ for the radius of 
gyration are independent of the knot type, the entropic exponent 
$\alpha$ for the number of accessible configurations clearly depends on 
the knot at hand. More precisely, $\alpha$ seems to depend only on the 
number of primary knots (factors) and increases by one for each added factor 
in the knot factorization.

In practice, all the characterizations of knots involve only their
projection in a two-dimensional (2D) plane. Therefore, nothing prevents us
from considering a {\it two-dimensional} model for knotted polymers. 
Such a model should be in principle much simpler to study, both
analytically and numerically. The aim of this paper is precisely to introduce 
and study numerically a particular model of ``projected" polygons on a 2D
lattice, describing 2D knotted polymers. In a 2D model for knotted polymers, 
the self-avoiding constraint in two dimensions is released to allow 
for what we call ``crossings". By crossings, we actually mean two
strands of the chain passing on top of each other in the projection. 
The object remains self-avoiding in the sense that the underlying and 
overlying strands cannot be exchanged by passing through each other. With 
this restriction, each closed polymer has a fixed knot type, which can be 
preserved in the deformations. Still the object is two-dimensional in the 
sense that its entropy corresponds to 2D deformations only. In particular,
we will recover in some limits some of the usual exponents of 2D 
self-avoiding walks (SAW).

The paper is organized as follows. In the first section, we introduce
our lattice model of projected self-avoiding polygons. Section 2 describes
the local elementary moves which we use to deform the polygon and explore
its accessible phase space at a fixed knot topology. The issue of 
ergodicity is also discussed in this section where arguments for 
a proof of ergodicity within a fixed knot type (but not a full proof) 
are given. The elementary moves are then performed according to a 
Monte Carlo Metropolis algorithm,
drawn to reproduce the most natural grand canonical statistical weight,
with a fugacity $K$ per bond of polygon, and a fugacity $w$ per
crossing.
The updating rules are discussed in section 3. Several improvements
of the algorithm are described in section 4. They involve both non local
deformations and multiple chains run in parallel. Our numerical
results are presented in sections 5, 6 and 7 for $w=1$, $w\to 0$
and $w>1$ respectively. We gather our conclusions in section 8.

\section{The Model}
\begin{figure}
\inseps{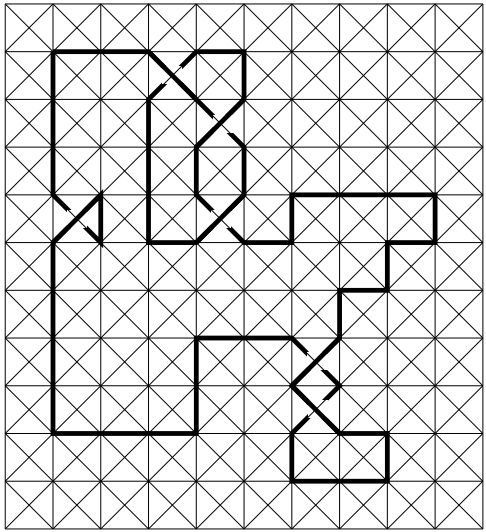}{0.5}
\caption{An example of a PSAP configuration on the square diagonal lattice 
with six crossings and the topology of a trefoil knot.}
\label{f:example}
\end{figure}
We present here the model of Projected Self-Avoiding Polygons (PSAP's) 
that we shall use to describe two-dimensional knotted polymers. The model is 
defined on the square diagonal lattice in two dimensions, namely the 
simple two dimensional square lattice $Z^2$ completed with the diagonals of 
the squares. We treat each diagonal as an elementary edge of the lattice, i.e.
we consider that the intersection point of the two diagonals is {\it not}
a vertex of the lattice. The bonds of the polymer can sit on all the edges 
of the lattice, either vertical, horizontal or diagonal, with at most one bond 
per lattice edge. Different rules apply for the edges and vertices of the 
simple square lattice on the one hand and for the diagonal edges on the other 
hand. We impose a strict self-avoidance at the vertices {\sl of the square 
lattice}, i.e. we do not allow two parts of the polymer to either cross each 
other or even to touch each other at these vertices. We allow crossings to 
take place inside a square {\sl at the intersection point of its two diagonal 
edges}. More precisely, we use the diagonal edges only for crossings, i.e. 
we impose that a diagonal edge can be occupied {\sl if and only if} the 
perpendicular diagonal edge in the same square is also occupied, with the two
bonds on these edges forming a crossing. For each pair of occupied diagonal 
edges inside a square, we distinguish between two different possible crossings 
according to which of the two bonds of polymer passes on top of the other. 
We can thus view crossings as a two dimensional projection of a pair of bonds 
in three dimensions, with one bond lying on top of the other. Finally, we 
limit ourselves to closed polymers. Figure \ref{f:example} shows an example 
of an allowed configuration with six crossings and the topology of a trefoil 
knot. 

Our model can be seen as a simple extension of the usual model describing
self-avoiding polymer loops on the square lattice, also refered to as
Self-Avoiding Polygons (SAP). The new ingredient here is the possibility for
the polymer, which is strictly self-avoiding on the square lattice, 
to have crossings taking place on pairs of perpendicular diagonal edges inside 
a square. Again, these ``crossings" are viewed as the two dimensional 
projection of two bonds of polymer passing on top of each other, with thus 
two distinct allowed configurations according to which of the two bonds is on 
top of the other. Thanks to this distinction, our model still describes a 
particular self-avoiding object in the sense that it mimics the
{\sl projection} in two dimensions of a polymer which would be self-avoiding 
in three dimensions. We will refer to our model as a model of 
Projected Self-Avoiding Polygons (PSAP's). Since we use closed polygons, 
a PSAP will in general form knots, i.e. will be the two-dimensional projection
of a three dimensional knot. In this sense, we can speak of our model
as describing two-dimensional knotted polymers. 

\begin{figure}
\inseps{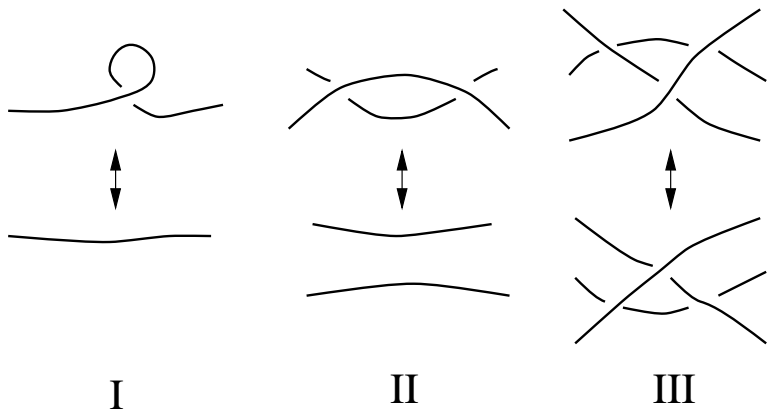}{0.5}
\caption{The three types (I, II and III) of Reidemeister moves
on the projections of knots. Reversing all the crossings also 
corresponds to allowed transformations.}
\label{f:reidemeister}
\end{figure}
To fully specify the model and to study the corresponding statistics, 
we need to assign to each PSAP configuration its weight. 
Since our aim is the study of knotted polymers of a fixed knot type, 
we want to attach a non-zero weight only to those configurations which 
have the desired knot-topology. We will be mainly interested below in rather
simple topologies, i.e. that of the unknot ($\emptyset$), that of the 
trefoil knot ($3_1$), that of the figure eight knot ($4_1$) and that of the 
composite knots made of two trefoils ($3_1\#3_1$) or a trefoil and a figure 
eight ($3_1\#4_1$). In practice, we will start with an initial configuration 
fixing the knot type and we will explore the phase space accessible by 
performing successive transformations (moves) on the PSAP which preserve 
its topology. As well known from knot theory, local deformations exist 
on the projection of a knot which preserve its topology. These
deformations are classified as the Reidemeister moves \cite{Adams}, which are 
of three types as described in figure \ref{f:reidemeister}. These 
topological deformations are sufficient to pass from any two configurations
of the same knot type. The set of allowed moves in our lattice model will 
be described in detail in the next section. For all the accessible 
conformations of PSAP of a given knot type, we moreover weigh the configuration 
by a factor 
\begin{equation}
\pi\left ({\cal{P}}\right ) = \frac{n^Q K^{n}w^{c}}{G_\tau(Q,K,w)}
\label{distrorig}
\end{equation}
where $n$ is the number of bonds of the PSAP configuration ${\cal P}$, 
with a fugacity $K$ per bond, and $c$ is the number of crossings,
with a fugacity $w$ per crossing. We also artificially introduce a factor 
$n^Q$ with a factor $Q>1$ for numerical convenience.
This factor is unimportant for averages at fixed $n$ (canonical ensemble)
but it can improve the statistics in a grand-canonical ensemble 
with varying $n$. We typically take $Q=2$ in the following.
The denominator in eq.(\ref{distrorig}) is the grand-canonical partition 
function at fixed knot-topology $\tau$, which normalizes the weights so 
that the total weight for all accessible configurations is equal to unity. 
For the knot-topology preserving moves, the above weight will dictate the 
probability to accept or reject the deformation.

\section{Local Elementary Moves}
In this section, we describe the local part of our algorithm, which 
is a {\sl grand-canonical} implementation, since it involves changes 
in the number of bonds of the projected polygon. Our algorithm uses 
local elementary moves of four different types depending on how many 
elementary squares of $Z^2$ (plaquettes) are involved in the move. 
We will use {\sl single}, {\sl double}, {\sl 3-plaquette} (or {\sl corner}) 
and {\sl 6-plaquette} moves.  We first give here a description of all these 
elementary moves. The way we decide in the algorithm which type of move we 
attempt and with which probability we accept or reject the attempted move 
is described in the next section. 

\begin{description}
\item[Single Moves]
Single Moves are performed on a single plaquette, and may 
involve length changes through the addition or subtraction of bonds
in the PSAP. This moves are defined as follows: given a 
horizontal or vertical bond $(p_i,p_{i+1})$ of the current PSAP, 
we pick a unit vector $e_i$ perpendicular to $(p_i,p_{i+1})$. This
defines a plaquette $(p_i,p_{i+1},p_{i+1}+e_i,p_i+e_i)$.
A move can occur only if the edge $(p_i+e_i,p_{i+1}+e_i)$ is not
occupied. The move depends on how many edges are occupied on the
plaquette. If exactly two edges are occupied, we exchange occupied
and unoccupied edges. This corresponds to a on bead flip (transformation
$S_I$ in figure \ref{f:BFACF}). If only one edge is occupied, we 
shift $(p_i,p_{i+1})$ one lattice spacing in the direction of $e_i$ and 
complete the PSAP by two additional bonds, either on the sides, creating a 
kink (transformation $S_{II}$ in figure \ref{f:BFACF}), or on the diagonals,
creating a crossing (transformation $S_{IV}$ or $S_{IV}'$
in figure \ref{f:BFACFcchanging}). If exactly three edges are occupied, 
we move $(p_i,p_{i+1})$
to $(p_{i}+e_i,p_{i+1}+e_i)$ and delete the two other bonds. The result
is either a kink deletion (transformation $S_{III}$ in figure \ref{f:BFACF})
or a crossing deletion (transformation $S_V$ or $S_V'$ in figure 
\ref{f:BFACFcchanging}).
The single moves can be classified into two different
sub-groups depending on whether they do or do not preserve the number of
crossings $c$ of the PSAP. 
The $c$-preserving moves illustrated in figure \ref{f:BFACF} are known
in the literature as the BFACF moves~\cite{BF81,AC83,ACF83}. 
While the move $S_I$ (one-bead flip) is $n$-preserving,
the moves $S_{II}$ (kink insertion) and $S_{III}$ (kink deletion) change 
the length of the PSAP respectively by $\Delta n = +2$ and $\Delta n = -2$.
The $c$-changing moves are illustrated in figure \ref{f:BFACFcchanging}. 
In this case, in addition to the change $\Delta n = \pm 2$ in the length of 
the PSAP, the number of crossings changes by an amount $\Delta c = + 1$ 
(moves $S_{IV}$ and $S_{IV}'$) or $\Delta c = - 1$ (moves $S_{V}$ 
and $S_{V}'$). The moves $S_{IV}'$ and $S_{V}'$ are identical to the moves 
$S_{IV}$ and $S_{V}$ apart from the reversing of the crossing between 
${\overline{\cal{S}}_4}$ and ${\cal{S}}_4$. All these $c$-changing moves
correspond to a Reidemeister I transformation.

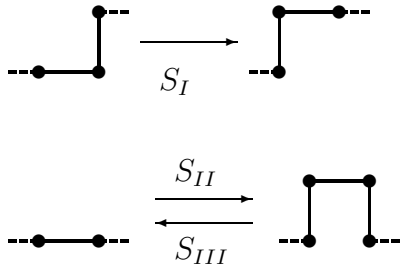
\begin{figure}
\unitlength = 8mm   
\newsavebox{\moveJK}
\savebox{\moveJK}(6,1){
 \begin{picture}(6,1)
  \thicklines
  \put(0,0){\usebox{\dashes}}   
  \put(0.5,0){\circle*{0.2}}   
  \drawline(0.5,0)(1.5,0)
  \put(1.5,0){\circle*{0.2}}   
  \put(1.5,0){\usebox{\dashes}}   
  \thinlines
  \put(2.45,0.7){\vector(1,0){1.6}}   
  \put(2.75,1){\bf $S_{II}$}   
  \put(4.05,0.3){\vector(-1,0){1.6}}  
  \put(2.75,-0.3){\bf $S_{III}$}   
  \thicklines
  \put(4.5,0){\usebox{\dashes}}   
  \put(5,0){\circle*{0.2}}   
  \drawline(5,0)(5,1)
  \put(5,1){\circle*{0.2}}   
  \drawline(5,1)(6,1)
  \put(6,1){\circle*{0.2}}   
  \drawline(6,1)(6,0)
  \put(6,0){\circle*{0.2}}   
  \put(6,0){\usebox{\dashes}}   
 \end{picture}
}
\newsavebox{\moveA}
\savebox{\moveA}(6,1){
 \begin{picture}(6,1)
  \thicklines
  \put(0,0){\usebox{\dashes}}   
  \put(0.5,0){\circle*{0.2}}   
  \drawline(0.5,0)(1.5,0)
  \put(1.5,0){\circle*{0.2}}   
  \drawline(1.5,0)(1.5,1)
  \put(1.5,1){\circle*{0.2}}   
  \put(1.5,1){\usebox{\dashes}}   
  \thinlines
  \put(2.2,0.5){\vector(1,0){1.6}}
  \thicklines
  \put(4,0){\usebox{\dashes}}   
  \put(4.5,0){\circle*{0.2}}   
  \drawline(4.5,0)(4.5,1)
  \put(4.5,1){\circle*{0.2}}   
  \drawline(4.5,1)(5.5,1)
  \put(5.5,1){\circle*{0.2}}   
  \put(5.5,1){\usebox{\dashes}}   
  \put(2.5,-0.3){\bf $S_{I}$}   
 \end{picture}
}
\begin{center}
\usebox{\moveA}
\end{center}
\vspace{.5cm}
\begin{center}
\usebox{\moveJK}
\end{center}
\caption{
   Single local $c$-preserving moves.
   ($S_{I}$) One-bead flip ($\Delta n = 0$).
   ($S_{II}$) Kink insertion ($\Delta n = +2$).
   ($S_{III}$) Kink deletion ($\Delta n = -2$).
}
\label{f:BFACF}
\end{figure}

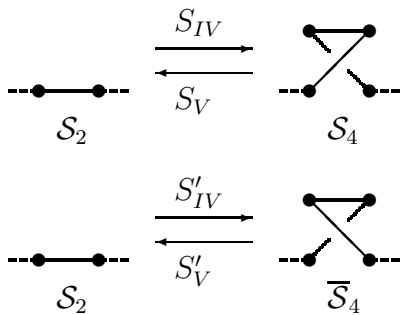
\begin{figure}
\unitlength = 8mm   
\newsavebox{\moveIV}
\savebox{\moveIV}(6,1){
 \begin{picture}(6,1)
  \thicklines
  \put(0,0){\usebox{\dashes}}   
  \put(0.5,0){\circle*{0.2}}   
  \drawline(0.5,0)(1.5,0)
  \put(1.5,0){\circle*{0.2}}   
  \put(1.5,0){\usebox{\dashes}}   
  \put(0.8,-0.8){${\cal{S}}_2$}   
  \thinlines
  \put(2.45,0.7){\vector(1,0){1.6}}   
  \put(2.75,1){\bf $S_{IV}$}   
  \put(4.05,0.3){\vector(-1,0){1.6}}  
  \put(2.75,-0.3){\bf $S_{V}$}   
  \thicklines
  \put(4.5,0){\usebox{\dashes}}   
  \put(5,0){\circle*{0.2}}   
  \drawline(5,0)(6,1)
  \put(5,1){\circle*{0.2}}   
  \drawline(5,1)(6,1)
  \put(6,1){\circle*{0.2}}   
  \drawline(5,1)(5.33,0.66)
  \drawline(5.66,0.33)(6,0)
  \put(6,0){\circle*{0.2}}   
  \put(6,0){\usebox{\dashes}}   
  \put(5.3,-0.8){${\cal{S}}_4$}   
 \end{picture}
}
\newsavebox{\moveV}
\savebox{\moveV}(6,1){
 \begin{picture}(6,1)
  \thicklines
  \put(0,0){\usebox{\dashes}}   
  \put(0.5,0){\circle*{0.2}}   
  \drawline(0.5,0)(1.5,0)
  \put(1.5,0){\circle*{0.2}}   
  \put(1.5,0){\usebox{\dashes}}   
   \put(0.8,-0.8){${\cal{S}}_2$}   
   \thinlines
  \put(2.45,0.7){\vector(1,0){1.6}}   
  \put(2.75,1){\bf $S_{IV}'$}   
  \put(4.05,0.3){\vector(-1,0){1.6}}  
  \put(2.75,-0.3){\bf $S_{V}'$}   
  \thicklines
  \put(4.5,0){\usebox{\dashes}}   
  \put(5,0){\circle*{0.2}}   
  \drawline(5,0)(5.33,0.33)
  \drawline(5.66,0.66)(6,1)
  \put(5,1){\circle*{0.2}}   
  \drawline(5,1)(6,1)
  \put(6,1){\circle*{0.2}}   
  \drawline(5,1)(6,0)
  \put(6,0){\circle*{0.2}}   
  \put(6,0){\usebox{\dashes}}   
  \put(5.3,-0.8){$\overline{\cal{S}}_4$}   
 \end{picture}
}

\begin{center}
\usebox{\moveIV}
\end{center}
\vspace{.5cm}
\begin{center}
\usebox{\moveV}
\end{center}
\caption{
   Single local $c$-changing moves: Reidemeister I moves.
   ($S_{IV}$,$S_{IV}'$) Crossing insertion ($\Delta c = +1$).
   ($S_{V}$,$S_{V}'$) Crossing deletion ($\Delta c = -1$).
}
\label{f:BFACFcchanging}
\end{figure}

\item[Double Moves]
A double move is performed on two adjacent plaquettes. Double moves are 
selected by first choosing, along the current PSAP, a vertex $p_i$. 
Different kinds of double moves are considered, depending on the relative 
orientations of the two bonds $(p_{i-1},p_i)$ and $(p_i,p_{i+1})$ shared by the 
vertex $p_i$ (see figure \ref{f:doublecases}):
\begin{description}
\item[Case of parallel vertical or horizontal bonds:] 
If the two bonds are parallel, we choose one of 
the two possible unit vectors $e_i$ perpendicular to $(p_{i-1},p_{i+1})$,
and we check if the vertex $p_j = p_i+e_i$ belongs to the current PSAP 
configuration. 
If it does, we look for a 2-plaquette configuration ${\cal{D}}_1$ such as that 
depicted in figure \ref{f:double1}. If this 2-plaquette configuration is
encountered, we make the local transformation $D_{I}$ by exchanging $p_i$
and $p_j=p_i+e_i$, leading to the configuration ${\cal{D}}_2$ with two more 
crossings, or make the similar transformation $D_{I}'$ leading to the
reversed configuration $\overline{\cal{D}}_2$.
\item[Case of perpendicular diagonal bonds:] 
If $(p_{i-1},p_i)$ and $(p_i,p_{i+1})$ are on perpendicular diagonal edges, the 
PSAP necessarily has two crossings involving two strands of polygon. We check 
that we have a configuration of type ${\cal{D}}_2$ or 
$\overline{\cal{D}}_2$ i.e. that the two strands are not entangled 
(see figure \ref{f:double1}). If so, we then make the local transformation 
$D_{II}$ or $D_{II}'$ to suppress the two crossings. 
\item[Case of bonds at $135^{\circ}$:]
If one of the bonds, say $(p_{i-1},p_i)$ is vertical or horizontal, and the other
bond $(p_i,p_{i+1})$ is at $135^{\circ}$ on a diagonal edge, the procedure is 
analogous to the case of parallel bonds, with $e_i$ the unit vector 
perpendicular to $(p_{i-1},p_{i})$ and inside the convex sector 
$(p_{i-1},p_i,p_{i+1})$. After checking that the vertex $p_j = p_i+e_i$ belongs 
to the current PSAP, we look for a 2-plaquette configuration such as the one
depicted in figure \ref{f:double2}. We then let the crossing diffuse
one step, according to transformation $D_{III}$ of figure \ref{f:double2}.
A similar procedure is used for the configuration $\overline{\cal{D}}_3$ 
obtained by reversing the crossing of ${\cal{D}}_3$.
\end{description}
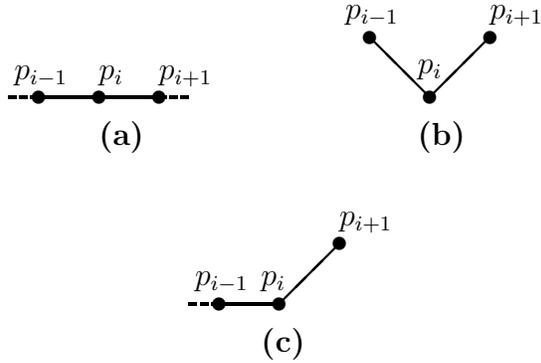
\begin{figure}
\unitlength = 8mm   
\newsavebox{\casedab}
\savebox{\casedab}(6,1){
 \begin{picture}(6,1)
  \thicklines
  \put(0,0){\usebox{\dashes}}   
  \put(0.5,0){\circle*{0.2}}   
  \put(0.1,0.3){$p_{i-1}$}   
  \drawline(0.5,0)(1.5,0)
  \put(1.5,0){\circle*{0.2}}   
  \put(1.5,0.3){$p_{i}$}   
  \drawline(1.5,0)(2.5,0)
  \put(2.5,0){\circle*{0.2}}   
  \put(2.5,0.3){$p_{i+1}$}   
  \put(2.5,0){\usebox{\dashes}}   
  \put(1.5,-0.8){\bf (a)}   
  \thicklines
  \put(6,1){\circle*{0.2}}   
  \put(5.6,1.3){$p_{i-1}$}   
  \drawline(6,1)(7,0)
  \put(7,0){\circle*{0.2}}   
  \put(6.8,0.4){$p_{i}$}   
  \drawline(7,0)(8,1)
  \put(8,1){\circle*{0.2}}   
  \put(8.,1.3){$p_{i+1}$}   
  \put(6.8,-0.8){\bf (b)}   
 \end{picture}
}
\newsavebox{\casedc}
\savebox{\casedc}(6,1){
 \begin{picture}(6,1)
  \thicklines
  \put(3,0){\usebox{\dashes}}   
  \put(3.5,0){\circle*{0.2}}   
  \put(3.1,0.3){$p_{i-1}$}   
  \drawline(3.5,0)(4.5,0)
  \put(4.5,0){\circle*{0.2}}   
  \put(4.2,0.3){$p_{i}$}   
  \drawline(4.5,0)(5.5,1)
  \put(5.5,1.0){\circle*{0.2}}   
  \put(5.5,1.3){$p_{i+1}$}   
  \put(4.2,-0.8){\bf (c)}   
  \thicklines
 \end{picture}
}
\begin{center}
\usebox{\casedab}
\end{center}
\vspace{1cm}
\begin{center}
\usebox{\casedc}
\end{center}
\caption{ Different cases considered for double moves: (a) parallel
vertical or horizontal bonds; (b) perpendicular diagonal bonds; (c)
bonds at $135^{\circ}$.
}
\label{f:doublecases}
\end{figure}
Finally, no double move is performed if the two consecutive bonds 
shared by $p_i$ belong to the same plaquette. In this case a corner 
move (see below) will be attempted instead.

Transformations $D_I$ and $D_{II}$ can be though of as Reidemeister II moves
while transformation $D_{III}$ is simply a diffusion of the crossing
to a neighboring plaquette.

In addition to the configurations described above, there are also 2-plaquette 
configurations in which the two involved strands of the PSAP are consecutive. 
We call these configurations {\sl degenerate} since can be transformed one into 
another by the same set of double moves above.
Some of such configurations are illustrated in figure \ref{f:doubledegenerate}.
\begin{figure}
\unitlength = 8mm   
\newsavebox{\moveda}
\savebox{\moveda}(6,1){
 \begin{picture}(6,1)
  \thicklines
  \put(0,0){\usebox{\dashes}}   
  \put(0.5,0){\circle*{0.2}}   
  \drawline(0.5,0)(1.5,0)
  \put(1.5,0){\circle*{0.2}}   
  \drawline(1.5,0)(2.5,0)
  \put(2.5,0){\circle*{0.2}}   
  \put(2.5,0){\usebox{\dashes}}   
  \put(0,1){\usebox{\dashes}}   
  \put(0.5,1){\circle*{0.2}}   
  \drawline(0.5,1)(1.5,1)
  \put(1.5,1){\circle*{0.2}}   
  \drawline(1.5,1)(2.5,1)
  \put(2.5,1){\circle*{0.2}}   
  \put(2.5,1){\usebox{\dashes}}   
  \put(0.8,-0.8){${\cal{D}}_1$}   
  \thinlines
  \put(3.45,0.7){\vector(1,0){1.6}}   
  \put(3.75,1){\bf $D_{I}$}   
  \put(5.05,0.3){\vector(-1,0){1.6}}  
  \put(3.75,-0.3){\bf $D_{II}$}   
  \thicklines
  \put(5.5,0){\usebox{\dashes}}   
  \put(5.5,1.0){\usebox{\dashes}}   
  \put(6,0){\circle*{0.2}}   
  \drawline(6,0)(7,1)
  \put(7,1){\circle*{0.2}}   
  \drawline(7,1)(8,0)
  \put(8,0){\circle*{0.2}}   
  \put(6,1){\circle*{0.2}}   
  \drawline(6,1)(6.33,0.66)
  \drawline(6.66,0.33)(7,0)
  \put(7,0){\circle*{0.2}}   
  \drawline(7,0)(7.33,0.33)
  \drawline(7.66,0.66)(8.0,1.0)
  \put(8,1.0){\circle*{0.2}}   
  \put(8,0){\usebox{\dashes}}   
  \put(8,1.0){\usebox{\dashes}}   
  \put(6.8,-0.8){${\cal{D}}_2$}   
 \end{picture}
}
\begin{center}
\usebox{\moveda}
\end{center}
\caption{
   Double $c$-changing moves : Reidemeister II moves.
}
\label{f:double1}
\end{figure}
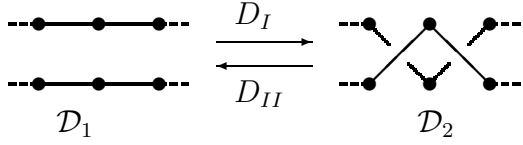
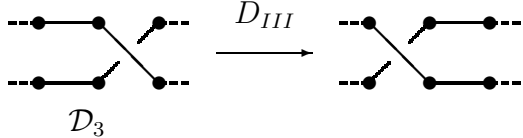
\begin{figure}
\unitlength = 8mm   
\newsavebox{\moveVII}
\savebox{\moveVII}(6,1){
 \begin{picture}(6,1)
  \thicklines
  \put(0,0){\usebox{\dashes}}   
  \put(0.5,0){\circle*{0.2}}   
  \drawline(0.5,0)(1.5,0)
  \put(1.5,0){\circle*{0.2}}   
  \drawline(1.5,0)(1.8333,0.333)
  \drawline(2.166,0.666)(2.5,1.0)
  \put(2.5,1.0){\circle*{0.2}}   
  \put(2.5,1.0){\usebox{\dashes}}   
  \put(0,1){\usebox{\dashes}}   
  \put(0.5,1){\circle*{0.2}}   
  \drawline(0.5,1)(1.5,1)
  \put(1.5,1){\circle*{0.2}}   
  \drawline(1.5,1)(2.5,0)
  \put(2.5,0){\circle*{0.2}}   
  \put(2.5,0){\usebox{\dashes}}   
  \put(1.0,-0.8){${\cal{D}}_3$}   
  \thinlines
  \put(3.45,0.5){\vector(1,0){1.6}}   
   \put(3.75,1){\bf $D_{III}$}   
  \thicklines
  \put(5.5,0){\usebox{\dashes}}   
  \put(5.5,1.0){\usebox{\dashes}}   
  \put(6,0){\circle*{0.2}}   
  \drawline(6,0)(6.33,0.33)
  \drawline(6.67,0.67)(7.,1.0)
  \put(7,1){\circle*{0.2}}   
  \drawline(7,1)(8,1)
  \put(8,1){\circle*{0.2}}   
  \put(8,1){\usebox{\dashes}}   
  \put(6,1){\circle*{0.2}}   
  \drawline(6,1)(7,0)
  \put(7,0){\circle*{0.2}}   
  \drawline(7,0)(8,0)
  \put(8,0){\circle*{0.2}}   
  \put(8,0){\usebox{\dashes}}   
 \end{picture}
}
\begin{center}
\usebox{\moveVII}
\end{center}
\caption{
   Double $c$-preserving move: diffusion of a crossing.
}
\label{f:double2}
\end{figure}
\begin{figure}
\unitlength = 8mm   
\newsavebox{\movedda}
\savebox{\movedda}(6,1){
 \begin{picture}(6,1)
  \thicklines
  \put(0.5,0){\circle*{0.2}}   
  \drawline(0.5,0)(1.5,0)
  \put(1.5,0){\circle*{0.2}}   
  \drawline(1.5,0)(2.5,0)
  \put(2.5,0){\circle*{0.2}}   
  \put(2.5,0){\usebox{\dashes}}   
  \drawline(0.5,0)(0.5,1)
  \put(0.5,1){\circle*{0.2}}   
  \drawline(0.5,1)(1.5,1)
  \put(1.5,1){\circle*{0.2}}   
  \drawline(1.5,1)(2.5,1)
  \put(2.5,1){\circle*{0.2}}   
  \put(2.5,1){\usebox{\dashes}}   
  \put(0.8,-0.8){${\cal{D}}_1$}   
  \thinlines
  \put(3.45,0.7){\vector(1,0){1.6}}   
  \put(3.75,1){\bf $D_{I}$}   
  \put(5.05,0.3){\vector(-1,0){1.6}}  
  \put(3.75,-0.3){\bf $D_{II}$}   
  \thicklines
  \drawline(6.0,0)(6,1)
  \put(6,0){\circle*{0.2}}   
  \drawline(6,0)(7,1)
  \put(7,1){\circle*{0.2}}   
  \drawline(7,1)(8,0)
  \put(8,0){\circle*{0.2}}   
  \put(6,1){\circle*{0.2}}   
  \drawline(6,1)(6.33,0.66)
  \drawline(6.66,0.33)(7,0)
  \put(7,0){\circle*{0.2}}   
  \drawline(7,0)(7.33,0.33)
  \drawline(7.66,0.66)(8.0,1.0)
  \put(8,1.0){\circle*{0.2}}   
  \put(8,0){\usebox{\dashes}}   
  \put(8,1.0){\usebox{\dashes}}   
  \put(6.8,-0.8){${\cal{D}}_2$}   
 \end{picture}
}
\newsavebox{\movedVII}
\savebox{\movedVII}(6,1){
 \begin{picture}(6,1)
  \thicklines
  \put(0.5,0){\circle*{0.2}}   
  \drawline(0.5,0)(1.5,0)
  \put(1.5,0){\circle*{0.2}}   
  \drawline(1.5,0)(1.8333,0.333)
  \drawline(2.166,0.666)(2.5,1.0)
  \put(2.5,1.0){\circle*{0.2}}   
  \put(2.5,1.0){\usebox{\dashes}}   
  \drawline(0.5,0)(0.5,1)
  \put(0.5,1){\circle*{0.2}}   
  \drawline(0.5,1)(1.5,1)
  \put(1.5,1){\circle*{0.2}}   
  \drawline(1.5,1)(2.5,0)
  \put(2.5,0){\circle*{0.2}}   
  \put(2.5,0){\usebox{\dashes}}   
  \put(1.0,-0.8){${\cal{D}}_3$}   
  \thinlines
   \put(3.45,0.5){\vector(1,0){1.6}}   
   \put(3.75,1){\bf $D_{III}$}   
  \thicklines
  \drawline(6,0)(6.,1)
  \put(6,0){\circle*{0.2}}   
  \drawline(6,0)(6.33,0.33)
  \drawline(6.67,0.67)(7.,1.0)
  \put(7,1){\circle*{0.2}}   
  \drawline(7,1)(8,1)
  \put(8,1){\circle*{0.2}}   
  \put(8,1){\usebox{\dashes}}   
  \put(6,1){\circle*{0.2}}   
  \drawline(6,1)(7,0)
  \put(7,0){\circle*{0.2}}   
  \drawline(7,0)(8,0)
  \put(8,0){\circle*{0.2}}   
  \put(8,0){\usebox{\dashes}}   
 \end{picture}
}
\begin{center}
\usebox{\movedda}
\end{center}
\vskip .8truecm
\begin{center}
\usebox{\movedVII}
\end{center}

\caption{
   Degenerate configurations for double moves.
}
\label{f:doubledegenerate}
\end{figure}
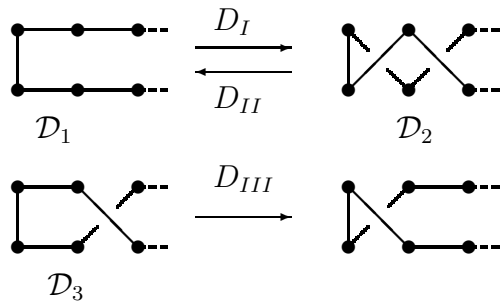
\item[Corner Moves]. 
In these moves, three plaquettes of the underlying squared lattice
are involved. As for the double moves, we first choose a vertex $p_i$
of the current PSAP.
A corner move can be performed if the bonds $(p_{i-1},p_i)$ and 
$(p_{i},p_{i+1})$, shared by the chosen vertex $p_i$, are on the two
consecutive edges of a square, as illustrated in figure \ref{f:cornercases}. 
In this case, we check if the vertex $p_j = -p_i + p_{i+1} + p_{i-1}$.
belongs to the current PSAP. If so, we look for local configurations involving 
three plaquettes such as those depicted in figure \ref{f:corner1} 
(${\cal{C}}_1$,${\cal{C}}_2$) and in figure \ref{f:corner2} (${\cal{C}}_3$). 
If the local 3-plaquette configuration does not correspond to any of such 
configurations no move is performed. Otherwise, we attempt one of the corner 
moves illustrated in figures \ref{f:corner1} and \ref{f:corner2}.  We can think 
of the corner moves $C_{I}$ and $C_{II}$ as Reidemeister II moves, whereas 
the corner move of figure \ref{f:corner2} corresponds to a diffusion
of a crossing through the corner. Note that this diffusion around a corner
also involves a rotation of the the crossing on the square lattice.
Corner moves $C_I'$ and $C_{II}'$ connecting
configuration ${\cal{C}}_1$ to $\overline{\cal{C}}_2$ are also possible.
In addition, as for the double moves, there are again degenerate 
configurations that can be transformed one into another by the
same corner moves described above. Some of them are illustrated in figure 
\ref{f:cornerdegenerate}.
\begin{figure}
\unitlength = 8mm   
\newsavebox{\casec}
\savebox{\casec}(6,1){
 \begin{picture}(6,1)
  \thicklines
  \put(2.9,-0.5){$p_{i-1}$}
  \put(3.5,0){\circle*{0.2}}   
  \drawline(3.5,0)(3.5,1)
  \put(4.3,-0.5){$p_{j}$}
  \put(4.3,-0.2){$\Diamond$}   
  \drawline(3.5,1)(4.5,1)
  \put(4.5,1.5){$p_{i+1}$}
  \put(4.5,1.){\circle*{0.2}}   
  \put(3.5,1){\circle*{0.2}}
   \put(3.1,1.5){$p_i$}
  \thicklines
 \end{picture}
}
\begin{center}
\usebox{\casec}
\end{center}
\caption{ Starting configuration for a corner move.}
\label{f:cornercases}
\end{figure}
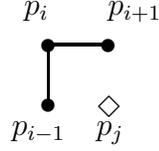
\begin{figure}
\unitlength = 8mm   
\newsavebox{\moveca}
\savebox{\moveca}(6,2){
 \begin{picture}(6,2)
  \thicklines
  \put(0.5,2){\circle*{0.2}}   
  \put(0.1,2.3){$p_{i}$}  
  \drawline(0.5,2)(1.5,2)
  \put(1.5,2){\circle*{0.2}}   
  \drawline(1.5,2)(2.5,2)
  \put(2.5,2){\circle*{0.2}}   
  \drawline(0.5,2)(0.5,1)
  \put(0.5,1){\circle*{0.2}}   
  \drawline(0.5,1)(0.5,0)
  \put(0.5,0){\circle*{0.2}}   
  \put(1.5,1){\circle*{0.2}}   
  \drawline(1.5,1)(2.5,1)
  \put(2.5,1){\circle*{0.2}}   
   \put(1.7,0.7){$p_j$}
  \drawline(1.5,1)(1.5,0)
  \put(1.5,0){\circle*{0.2}}   
  \put(0.5,-0.8){${\cal{C}}_1$}   
  \thinlines
  \put(3.45,1.3){\vector(2,0){1.6}}   
  \put(3.75,0.5){\bf $C_{I}$}   
  \put(5.05,1.7){\vector(-1,0){1.6}}  
  \put(3.75,2.3){\bf $C_{II}$}   
  \thicklines
  \put(5.6,2.3){$p_{i}$}  
  \put(6,2){\circle*{0.2}}   
  \drawline(6,2)(7,2)
  \put(7,2){\circle*{0.2}}   
  \drawline(7,2)(7.33,1.67)
  \drawline(7.67,1.33)(8.0,1.0)
  \put(8,1.0){\circle*{0.2}}   
  \drawline(6,1)(6,2)
  \put(6,1){\circle*{0.2}}   
  \drawline(6,1)(6.33,0.67)
  \drawline(6.67,0.33)(7,0)
  \put(7,0){\circle*{0.2}}   
  \put(6,0){\circle*{0.2}}   
  \drawline(6,0)(7,1)
  \put(7,1){\circle*{0.2}}   
  \put(7.2,0.8){$p_j$}  
  \drawline(7,1)(8,2)
  \put(8,2){\circle*{0.2}}   
  \put(6.2,-0.8){${\cal{C}}_2$}   
 \end{picture}
}
\begin{center}
\usebox{\moveca}
\end{center}
\caption{
   Corner $c$-changing move: Reidemeister II move around
    a corner.
}
\label{f:corner1}
\end{figure}
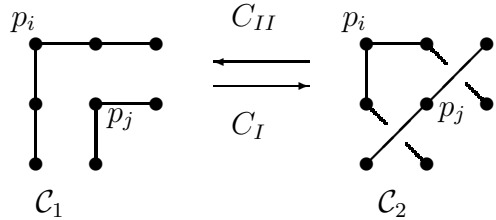
\begin{figure}
\unitlength = 8mm   
\newsavebox{\movecb}
\savebox{\movecb}(6,2){
 \begin{picture}(6,2)
  \thicklines
  \put(0.5,2){\circle*{0.2}}   
  \put(0.1,2.3){$p_{i}$}  
  \drawline(0.5,2)(1.5,2)
  \put(1.5,2){\circle*{0.2}}   
  \put(1.7,0.7){$p_j$}
  \drawline(1.5,2.)(1.83,1.67)
  \drawline(2.17,1.33)(2.5,1.)
  \put(2.5,1){\circle*{0.2}}   
  \drawline(0.5,2)(0.5,1)
  \put(0.5,1){\circle*{0.2}}   
  \drawline(0.5,1)(0.5,0)
  \put(0.5,0){\circle*{0.2}}   
  \put(1.5,1){\circle*{0.2}}   
  \drawline(1.5,1)(2.5,2.)
  \put(2.5,2){\circle*{0.2}}   
  \drawline(1.5,1)(1.5,0)
  \put(1.5,0){\circle*{0.2}}   
  \put(0.5,-0.8){${\cal{C}}_3$}   
  \thinlines
  \put(3.45,1.3){\vector(2,0){1.6}}   
  \put(3.75,0.5){\bf $C_{III}$}   
  \thicklines
  \put(5.6,2.3){$p_{i}$}
  \put(6,2){\circle*{0.2}}   
  \drawline(6,2)(7,2)
  \put(7,2){\circle*{0.2}}   
  \drawline(7,2)(8,2.)
  \put(8,2.0){\circle*{0.2}}   
  \drawline(6,1)(6,2)
  \put(6,1){\circle*{0.2}}   
  \drawline(6,1)(7,0.)
  \put(7,0){\circle*{0.2}}   
  \put(6,0){\circle*{0.2}}   
  \drawline(6,0)(6.33,0.33)
  \drawline(6.67,0.67)(7.,1.0)
  \put(7,1){\circle*{0.2}}   
  \put(7.15,0.6){$p_j$}
  \drawline(7,1)(8,1)
  \put(8,1){\circle*{0.2}}   
 \end{picture}
}
\begin{center}
\usebox{\movecb}
\end{center}
\caption{
   Corner $c$-preserving move : diffusion of a crossing around 
    a corner.
}
\label{f:corner2}
\end{figure}
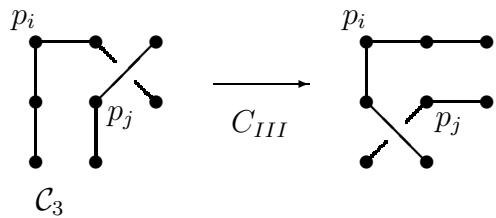

\begin{figure}
\unitlength = 8mm   
\newsavebox{\movedca}
\savebox{\movedca}(6,2){
 \begin{picture}(6,2)
  \thicklines
  \put(0.5,2){\circle*{0.2}}   
  \put(0.1,2.3){$p_{i}$}  
  \drawline(0.5,2)(1.5,2)
  \put(1.5,2){\circle*{0.2}}   
  \drawline(1.5,2)(2.5,2)
  \put(2.5,2){\circle*{0.2}}   
  \drawline(0.5,2)(0.5,1)
  \put(0.5,1){\circle*{0.2}}   
  \drawline(0.5,1)(0.5,0)
  \put(0.5,0){\circle*{0.2}}   
  \drawline(0.5,0)(1.5,0)

  \put(1.5,1){\circle*{0.2}}   
  \drawline(1.5,1)(2.5,1)
  \put(2.5,1){\circle*{0.2}}   
   \put(1.7,0.7){$p_j$}
  \drawline(1.5,1)(1.5,0)
  \put(1.5,0){\circle*{0.2}}   
  \put(0.5,-0.8){${\cal{C}}_1$}   
  \thinlines
  \put(3.45,1.3){\vector(2,0){1.6}}   
  \put(3.75,0.5){\bf $C_{I}$}   
  \put(5.05,1.7){\vector(-1,0){1.6}}  
  \put(3.75,2.3){\bf $C_{II}$}   
  \thicklines
  \put(5.6,2.3){$p_{i}$}  
  \put(6,2){\circle*{0.2}}   
  \drawline(6,2)(7,2)
  \put(7,2){\circle*{0.2}}   
  \drawline(7,2)(7.33,1.67)
  \drawline(7.67,1.33)(8.0,1.0)
  \put(8,1.0){\circle*{0.2}}   
  \drawline(6,1)(6,2)
  \put(6,1){\circle*{0.2}}   
  \drawline(6,1)(6.33,0.67)
  \drawline(6.67,0.33)(7,0)
  \put(7,0){\circle*{0.2}}   
  \put(6,0){\circle*{0.2}}   
  \drawline(6,0)(7,0)
  \drawline(6,0)(7,1)
  \put(7,1){\circle*{0.2}}   
  \put(7.2,0.8){$p_j$}  
  \drawline(7,1)(8,2)
  \put(8,2){\circle*{0.2}}   
  \put(6.2,-0.8){${\cal{C}}_2$}   
 \end{picture}
}
\newsavebox{\movedcb}
\savebox{\movedcb}(6,2){
 \begin{picture}(6,2)
  \thicklines
  \put(0.5,2){\circle*{0.2}}   
  \put(0.1,2.3){$p_{i}$}  
  \drawline(0.5,2)(1.5,2)
  \put(1.5,2){\circle*{0.2}}   
  \put(1.7,0.7){$p_j$}
  \drawline(1.5,2.)(1.83,1.67)
  \drawline(2.17,1.33)(2.5,1.)
  \put(2.5,1){\circle*{0.2}}   
  \drawline(0.5,2)(0.5,1)
  \put(0.5,1){\circle*{0.2}}   
  \drawline(0.5,1)(0.5,0)
  \put(0.5,0){\circle*{0.2}}   
  \put(1.5,1){\circle*{0.2}}   
  \drawline(1.5,1)(2.5,2.)
  \put(2.5,2){\circle*{0.2}}   
  \drawline(1.5,1)(1.5,0)
  \put(1.5,0){\circle*{0.2}}   
  \drawline(0.5,0)(1.5,0)
  \put(0.5,-0.8){${\cal{C}}_3$}   
  \thinlines
  \put(3.45,1.3){\vector(2,0){1.6}}   
  \put(3.75,0.5){\bf $C_{III}$}   
  \thicklines
  \put(5.6,2.3){$p_{i}$}
  \put(6,2){\circle*{0.2}}   
  \drawline(6,2)(7,2)
  \put(7,2){\circle*{0.2}}   
  \drawline(7,2)(8,2.)
  \put(8,2.0){\circle*{0.2}}   
  \drawline(6,0)(7,0)
  \drawline(6,1)(6,2)
  \put(6,1){\circle*{0.2}}   
  \drawline(6,1)(7,0.)
  \put(7,0){\circle*{0.2}}   
  \put(6,0){\circle*{0.2}}   
  \drawline(6,0)(6.33,0.33)
  \drawline(6.67,0.67)(7.,1.0)
  \put(7,1){\circle*{0.2}}   
  \put(7.15,0.6){$p_j$}
  \drawline(7,1)(8,1)
  \put(8,1){\circle*{0.2}}   
 \end{picture}
}
\begin{center}
\usebox{\movedca}
\end{center}
\vskip 0.8truecm
\begin{center}
\usebox{\movedcb}
\end{center} 
\caption{Degenerate cases of corner moves.}
\label{f:cornerdegenerate}
\end{figure}
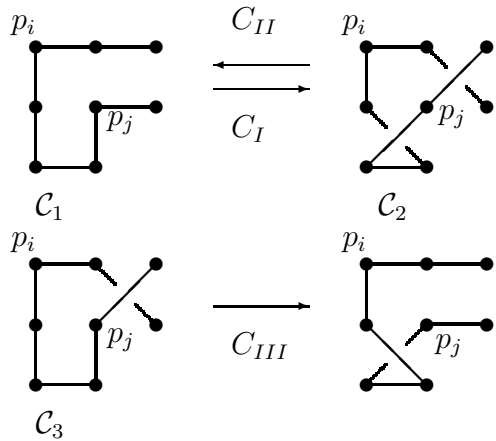
\item[6-plaquette Moves] (Reidemeister III moves).
This moves are performed by choosing a diagonal bond in the current PSAP. 
This bond singles out a crossing. We then look
for the presence of two neighboring crossings along two {\it adjacent}
diagonal directions, chosen at random. We finally look for
configurations such as the one depicted in figure \ref{f:RIII}. 
The performed move $R_I$ corresponds to a Reidemeister III move. 
Here again degenerate configurations exist which can be changed one into another.
\end{description}
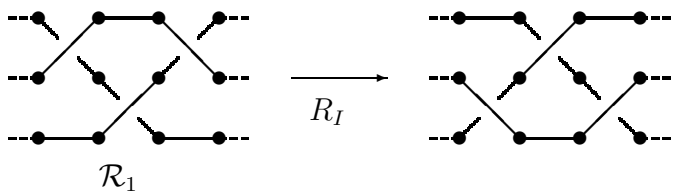
\begin{figure}
\unitlength = 8mm   
\newsavebox{\moveRIII}
\savebox{\moveRIII}(6,1){
 \begin{picture}(6,1)
  \thicklines
  \put(-1.0,0){\usebox{\dashes}}   
  \put(-0.5,0){\circle*{0.2}}   
  \drawline(-0.5,0)(0.5,0.)
  \put(0.5,0){\circle*{0.2}}   
  \drawline(0.5,0)(1.5,1)
  \put(1.5,1){\circle*{0.2}}   
  \drawline(1.5,1)(1.83,1.33)
  \drawline(2.17,1.67)(2.5,2.0)
  \put(2.5,2){\circle*{0.2}}   
  \put(2.5,2){\usebox{\dashes}}   
  \put(-1.0,1){\usebox{\dashes}}   
  \put(-0.5,1){\circle*{0.2}}   
  \drawline(-0.5,1)(0.5,2)
  \put(0.5,2){\circle*{0.2}}   
  \drawline(0.5,2)(1.5,2)
  \put(1.5,2){\circle*{0.2}}   
  \drawline(1.5,2)(2.5,1)
  \put(2.5,1){\circle*{0.2}}   
  \put(2.5,1){\usebox{\dashes}}   

  \put(-1.0,2){\usebox{\dashes}}   
  \put(-0.5,2){\circle*{0.2}}   
  \drawline(-0.5,2)(-0.17,1.67)
  \drawline(0.17,1.33)(0.5,1.)
  \put(0.5,1.0){\circle*{0.2}}   
  \drawline(0.5,1.0)(0.83,0.67)
  \drawline(1.17,0.33)(1.5,0)
  \put(1.5,0){\circle*{0.2}}   
  \drawline(1.5,0)(2.5,0)
  \put(2.5,0){\circle*{0.2}}   
  \put(2.5,0){\usebox{\dashes}}   

  \put(0.5,-0.8){${\cal{R}}_1$}   
  \thinlines
  \put(3.7,1.0){\vector(2,0){1.6}}   
  \put(4.0,0.3){\bf $R_I$}   
%
  \thicklines
  \put(6.0,0){\usebox{\dashes}}   
  \put(6.5,0){\circle*{0.2}}   
  \drawline(6.5,0)(6.83,0.33)
  \drawline(7.17,0.67)(7.5,1.0)
  \put(7.5,1.0){\circle*{0.2}}   
  \drawline(7.5,1.0)(8.5,2)
  \put(8.5,2){\circle*{0.2}}   
  \drawline(8.5,2)(9.5,2.)
  \put(9.5,2){\circle*{0.2}}   
  \put(9.5,2){\usebox{\dashes}}   
  \put(6.0,1){\usebox{\dashes}}   
  \put(6.5,1){\circle*{0.2}}   
  \drawline(6.5,1)(7.5,0)
  \put(7.5,0){\circle*{0.2}}   
  \drawline(7.5,0)(8.5,0)
  \put(8.5,0){\circle*{0.2}}   
  \drawline(8.5,0)(9.5,1)
  \put(9.5,1){\circle*{0.2}}   
  \put(9.5,1){\usebox{\dashes}}   

  \put(6.0,2){\usebox{\dashes}}   
  \put(6.5,2){\circle*{0.2}}   
  \drawline(6.5,2)(7.5,2)
  \put(7.5,2.0){\circle*{0.2}}   
  \drawline(7.5,2.0)(7.83,1.67)
  \drawline(8.17,1.33)(8.5,1)
  \put(8.5,1){\circle*{0.2}}   
  \drawline(8.5,1)(8.83,0.67)
  \drawline(9.17,0.33)(9.5,0)
  \put(9.5,0){\circle*{0.2}}   
  \put(9.5,0){\usebox{\dashes}}   
 \end{picture}
}
\begin{center}
\usebox{\moveRIII}
\end{center}
\caption{
   6-plaquette $c$-preserving move : Reidemeister III move.
}
\label{f:RIII}
\end{figure}
Note that double, corner and 6-plaquette moves do not change
the number $n$ of bonds in the PSAP.
\vskip 0.5cm
To end this section let us mention that:

\begin{itemize}
\item For all the moves drawn above, equivalent moves
exist which are obtained by $90^{\circ}$ rotations or mirror
reflections.
\item If after our checks we do not find one of the above
described environments, no move is performed.
\item As described in the next section we have an exhaustive
procedure to decide which type of move will be attempted.
\end{itemize}

All the moves above clearly preserve the knot-topology. Therefore any possible 
deformation will maintain the PSAP inside the set of conformations having
the same knot type as the conformation one started with. In the following, 
we will assume that the algorithm is {\it ergodic}, although a full proof of 
ergodicity goes far beyond our goal. 
By ergodicity, we mean that starting from a given conformation, any other 
conformation with the same knot-topology {\it can be obtained} by a series of 
successive elementary moves. In the absence of crossings, our algorithm 
reduces to the BFACF algorithm~\cite{BF81,AC83,ACF83}, which is known to be 
ergodic. More precisely, the BFACF moves (moves $S_I$, $S_{II}$ and
$S_{III}$) allow to deform any configuration of a set of self-avoiding
paths onto any other configuration with the same topology and, if some of the
paths are open, with the same positions of the end-points. Our algorithm 
completes the BFACF algorithm with knot-topology preserving moves allowing 
the three types of Reidemeister moves, as well as the migration of a crossing 
along the PSAP. 

\begin{figure}
\inseps{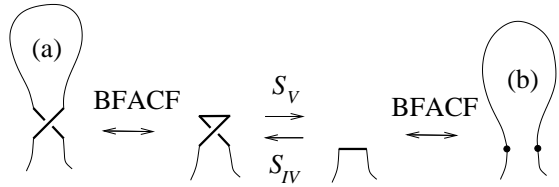}{0.4}
\caption{The passage between two expanded configurations (a) and (b) 
related by a Reidemeister I transformation. Thin lines indicate
an arbitrary (supposedly drawn a the square lattice) path joining 
the crossings. For each step of the transformation, we indicate 
the type of elementary move involved.
}
\label{f:moveI}
\end{figure}

\begin{figure}
\inseps{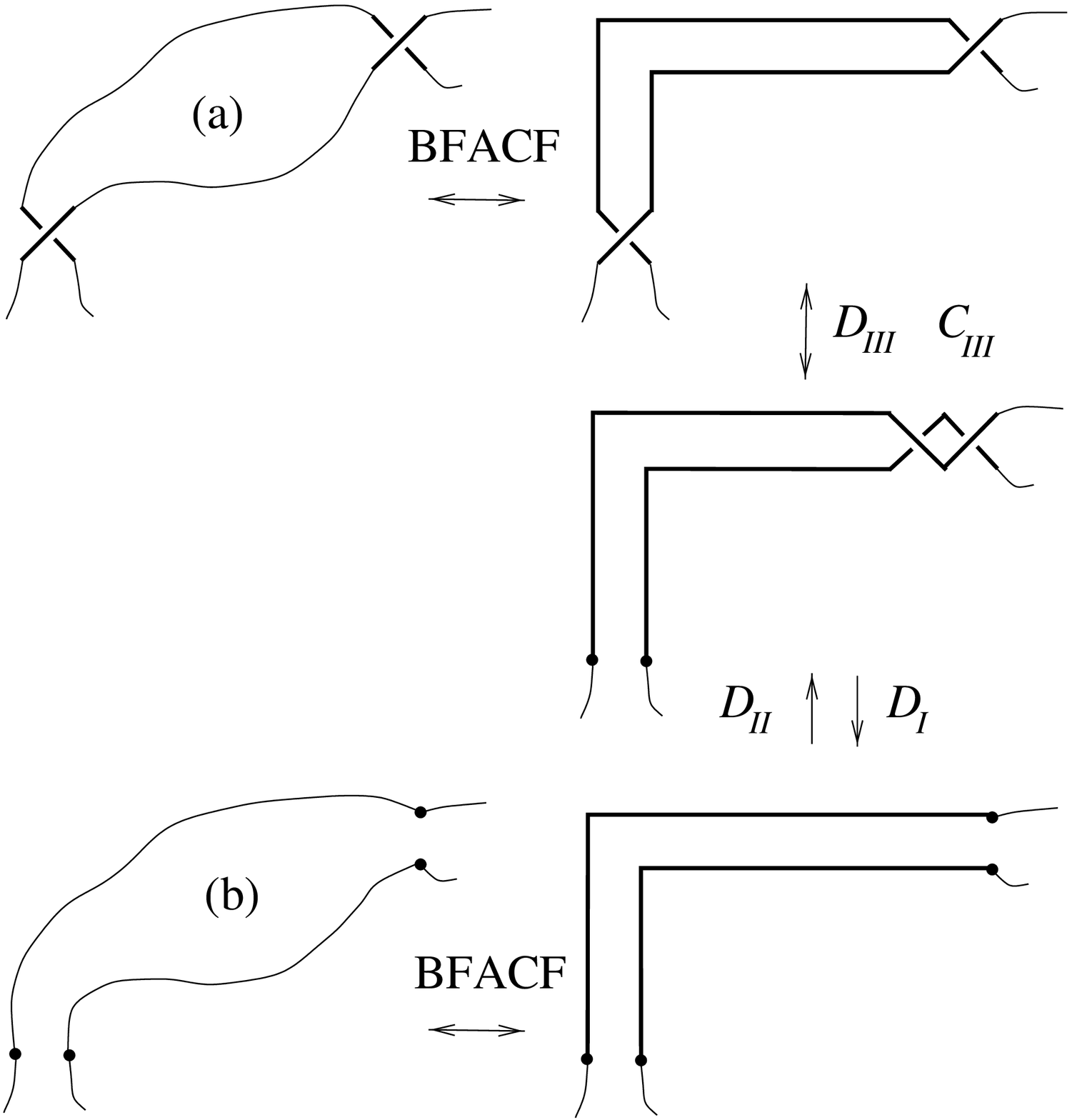}{0.4}
\caption{The passage between two expanded configurations (a) and (b)
related by a Reidemeister II transformation.}
\label{f:moveII}
\end{figure}

\begin{figure}
\inseps{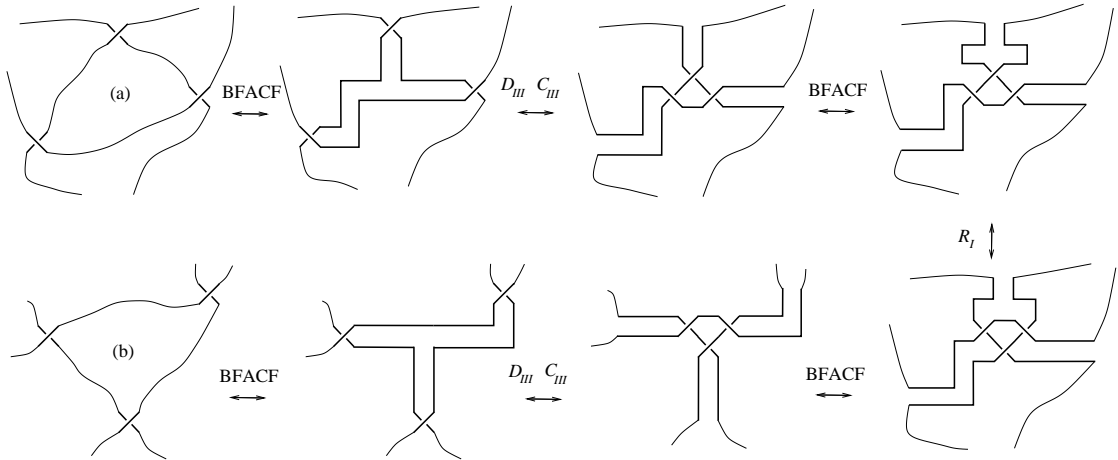}{.8}
\caption{The passage between two expanded configurations (a) and (b)
related by a Reidemeister III transformation.}
\label{f:moveIII}
\end{figure}

\begin{figure}
\inseps{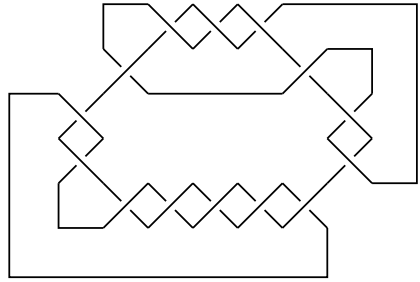}{.3}
\caption{A blocked configuration with a loop of 14 alternating crossings.}
\label{f:loop}
\end{figure}

Although we have no rigorous proof of ergodicity for our algorithm, let
us present here what could be the sketch of a proof. Before we proceed,
let us note that we can limit ourselves without loss of generality to 
conformations on which none of the transformations $S_{V}$, $C_{II}$
or $D_{II}$ 
(which decrease the number of crossings) can be performed. If such a 
transformation can be performed, then we do perform it (possibly recursively)
so as to eliminate the spurious crossings. We are thus left with 
configurations where the suppression of a crossing requires more than one 
elementary move. We then divide the set of remaining conformations 
with a fixed knot type into subsets of conformations which have precisely the 
{\it same} (topologically speaking) knot projection. The topology
of such a subset is characterized by a fixed number of crossings, and
a fixed prescription for the connectivity of these crossings.
A proof of ergodicity can then be established in two steps:
\begin{itemize}
\item{(1)} Proving the ergodicity within a subset, i.e. for a fixed topology
of the projection (i.e. a fixed number of crossings and a fixed set
of connections between them).
\item{(2)} Showing the possibility to pass from a {\it particular} element 
(representative) of one subset to a particular element of another subset if
the two different projections represent the same knot.
\end{itemize}
Since two projections representing the same knot can be related
by a series of Reidemeister moves, it is enough to prove point (2)
that representatives of two different subsets differing by a Reidemeister move
can be related one to the other. A good choice of representatives is what 
could be called expanded configurations, i.e. configurations in which 
all the crossings are well separated. In a grand-canonical ensemble,
such configurations always exist, whatever big is the number of crossings
in the projection. A Reidemeister move will involve at most three
crossings in the projection. For well separated crossings, the
figures \ref{f:moveI}, \ref{f:moveII} and \ref{f:moveIII} show
how to perform a Reidemeister transformation of type I, II or III
respectively by use of our elementary moves. For a Reidemeister I move, 
we can use the BFACF moves to shrink a path with neighboring end-points
onto the segment joining these end-points, then use the move $S_{IV}$
or $S_{V}$ to create or destroy a crossing, and use BFACF moves
again to re-inflate the path. For Reidemeister moves II and III,
the idea is to use the BFACF moves to deform the paths (with fixed 
end-points) connecting crossings so as to create channels along which 
the crossings can migrate. The migration of crossings along a channel 
is performed by use of the elementary moves $D_{III}$ or $C_{III}$ (at
a corner) until the crossings come in contact and an elementary move 
$D_{I}$ or $D_{II}$ (Reidemeister II), or $R_{I}$ (Reidemeister III)
can be performed. The target expanded configuration can then be reached
again by creating channels, moving the crossings to their final
positions, and eventually re-deforming the connecting paths.
Note that for all these deformations, the moves $C_{I}$ and $C_{II}$
have not been used. This is because these moves are not independent
and can be performed by a sequence of $D_{I}$, $D_{II}$, $D_{III}$, 
$C_{III}$ and BFACF moves. Still in a Monte Carlo program, it is
worth implementing them to accelerate the algorithm.
\vglue .5cm
Let us now come to the point (1) of our argument, i.e the ergodicity
of the algorithm inside the subset of conformations having the same
projection. A topological deformation will involve now only a displacement
of the moves and a deformation of the paths connecting them, keeping
the global connectivity structure. No Reidemeister move and in particular
no creation of crossings have to be performed. Since BFACF moves allow
to freely deform the already existing (i.e. with non-zero initial length)
paths, and the moves $D_{III}$ and $C_{III}$ allow to move any crossing 
along a channel of connecting paths, the only  possible source of blocking 
is due to situations where crossings have zero length connecting paths, i.e. 
are nearest neighbors of other crossings, without path between them. 
Indeed, our moves do not allow us to move several crossings at once. 
For this situation to lead to a real blocking, one moreover needs a set 
of neighboring crossings whose centers form a loop on the dual square lattice
(the centers of the squares), so that each crossing in the loop in blocked on 
both sides (see an example in figure \ref{f:loop}). Since spurious crossings 
have already been removed, these loop conformations require moreover many 
{\it alternating} crossings (at least eight) and correspond to entangled 
situations with a complex knot topology. 
We do not expect such conformations to belong to the simple knot topologies
that we study here where we consider knots made only of trefoils and figure eight 
primary knots.

\section{Updating and Reversibility}
We will now discuss the updating of our system, i.e. 
how we decide which type of move to attempt and with which probability
we accept the move. These choices are dictated by a condition of reversibility 
of the algorithm and by the nature of the invariant limiting distribution 
which we want to reach within an ergodicity class, i.e. the weight
$\pi({\cal P})$ of eq.(\ref{distrorig}) which we assign to a 
configuration $\cal P$ of the PSAP.

The implementation of the algorithm is ``Metropolis style". Let ${\cal{P}}_1$ 
be the current PSAP with $n_1$ bonds and $c_1$ crossings. A tentative move
of type $M$ is chosen according to a procedure which we describe below. 
It may or may not be possible. If it is, it produces a new configuration 
${\cal{P}}_2$ with 
length $n_2$ and number of crossings $c_2$. Then we accept the move with the 
acceptance rate 
\begin{equation}
{\rm Acc}( {\cal{P}}_1 \rightarrow {\cal{P}}_2 ) = 
A(M)\min \left[ 1, K^{n_2-n_1} \left(\frac{n_2}{n_1}\right)^{Q-1}
\right] \min\left[1, w^{c_2-c_1}\right]
\end{equation}
The two $\min[\cdot]$ terms correspond to the usual Metropolis acceptance 
rate dictated by the limiting distribution of eq. (\ref{distrorig}).
In particular, the term involving the ratio $n_2/n_1$ is necessary
to reproduce the $n^Q$ dependence of $\pi({\cal P})$.
In addition, we also introduce a coefficient $A(M)\le 1$ depending on the 
type of move $M$ attempted, and which we will determine later. 
The probability to pass from the configuration ${\cal P}_1$ to the configuration 
${\cal P}_2$ is thus 
\begin{equation}
\Prob( {\cal{P}}_1 \rightarrow {\cal{P}}_2 ) = 
A(M)\min \left[1, K^{n_2-n_1} \left(\frac{n_2}{n_1}\right)^{Q-1}
\right] \min\left[1,w^{c_2-c_1}\right]
\times B({\cal P}_1,M)
\end{equation}
where $B({\cal P}_1,M)$ is the probability to have selected the move $M$ among
all the tentative moves which could have been tried on ${\cal P}_1$.
To be sure that the Markov chain converges to the limiting distribution 
$\pi\left ({\cal{P}}\right )$, we impose
on $\Prob( {\cal{P}}_1 \rightarrow {\cal{P}}_2 )$ the detailed balance 
condition
\begin{equation}
\pi\left ({\cal{P}}_1\right ) \Prob\{ {\cal{P}}_1 \rightarrow {\cal{P}}_2 \}
= 
\pi\left ({\cal{P}}_2\right ) \Prob\{ {\cal{P}}_2 \rightarrow {\cal{P}}_1 \}
\label{detbal}
\end{equation}
This will dictate the choice of $A(M)$ below. 
\vskip 0.5cm
The implementation of the algorithm 
is as follows:

\begin{enumerate}

\item To attempt a local move on the current configuration ${\cal P}_1$
we first select with probability $q$ a bond at random on the configuration
or with probability $(1-q)$ a vertex at random on the configuration.
The number $q$ is a free parameter which we can adjust in our simulation.

\item Case of bond selection. In this case, after having chosen 
the bond at random among the $n_1$ bonds of ${\cal P}_1$, we attempt
a $6$-plaquettes move if the bond is a diagonal bond, or a
single move if the bond is vertical or horizontal.
\begin{description}
\item[(i)] 6-plaquette move. Starting from the chosen diagonal bond, 
we look for a configuration such as those in figure \ref{f:RIII} (or their 
degenerate companions). If such a
configuration is not found the move is rejected. Otherwise, the outcome
is a PSAP ${\cal{P}}_2$ with $n_2 = n_1$ and $c_2 = c_1$, and
\begin{equation}
\Prob\{ {\cal{P}}_1 \rightarrow {\cal{P}}_2 \} = A(R_I) \  B({\cal P}_1,R_I)
\quad ; \quad B({\cal P}_1,R_I)= \frac{2q}{n_1}
\times \frac{1}{4},
\end{equation}
The factor $2/n_1$ is the probability to have chosen among $n_1$
bonds one of the two bonds forming the crossing which is the
starting point of the Reidemeister III move, and the factor $1/4$ is
the probability to have selected the correct diagonal directions of
the two other crossings involved. The important point is that 
every $6$-plaquette move has a unique inverse (see figure \ref{f:RIII}) 
by design which is also a 6-plaquette move of the same type. 
Detailed balance holds by choosing
\begin{equation}
A(R_I)=1
\end{equation}
since, by exchanging ${\cal P}_1$ and ${\cal P}_2$, 
and using $n_2=n_1$, we have 
$\Prob\{ {\cal{P}}_2 \rightarrow {\cal{P}}_1 \}=
\Prob\{ {\cal{P}}_1 \rightarrow {\cal{P}}_2 \}$
and $\pi({\cal P}_1)=\pi({\cal P}_2)$.

\item[(ii)] Single move.
For the single moves, we proceed as described in the previous section.
If the considered plaquette has two occupied edges, we do a bead flip $S_I$.
Since the bead flip $S_I$ keeps the number of bonds and crossings unchanged,
the detailed balance follows by noticing that any such move has its own inverse 
by design. We can thus take
\begin{equation}
A(S_I)=1.
\end{equation}
If the considered plaquette has a single occupied edge, we must choose between
the three moves corresponding to an addition of two bonds ($S_{II}$,$S_{IV}$ 
and  $S_{IV}'$) with $n_2-n_1= 2$ and $\Delta c = 0$ or $1$. We choose at 
random one of
these three moves with even probability $1/3$. For convenience, the acceptance 
rates $A(S_{II})$, $A(S_{IV})$ and $A(S_{IV}')$ will be taken all equal to 
$A^{(S)}_{+2}$.
To fix them, we need to consider the reverse moves, which are performed on 
a plaquette which has three occupied edges. In this case, the current
configuration dictates without ambiguity which transformation $S_{III}$, 
$S_{V}$ or $S_{V}'$ can be done. Again we take $A(S_{III})$, $A(S_{V})$ and $A(S_{V}')$
all equal to $A^{(S)}_{-2}$. If ${\cal P}_1$ and 
${\cal P}_2$ are the two configurations 
mutually exchanged by some of these transformations, with $n_2=n_1+2$, we have
\begin{equation}
\Prob\{{\cal P}_1\rightarrow{\cal P}_2\}=A^{(S)}_{+2}\frac {q}{n_1}
\frac{1}{2}\,\frac{1}{3}\min\kern-4pt\left[1,K^{n_2-n_1}\left(\frac{n_2}{n_1}
\right)^{Q-1}\right]\min\kern-4pt\left[1,w^{c_2-c_1} \right]
\end{equation}
while
\begin{equation}
\Prob\{{\cal P}_2\rightarrow{\cal P}_1\}=A^{(S)}_{-2}\frac {q}{n_2}
\frac{1}{2}\min\kern-4pt\left[1,K^{n_1-n_2}\left(\frac{n_1}{n_2}
\right)^{Q-1}\right]\min\kern-4pt\left[1,w^{c_1-c_2}\right]
\end{equation}
Choosing 
\begin{equation}
A^{(S)}_{+2}=1 ; \quad A^{(S)}_{-2}=\frac{1}{3}
\end{equation}
the detailed balance
\begin{equation}
n_1^Q K^{n1}w^{c_1}\Prob\{ {\cal{P}}_1 \rightarrow {\cal{P}}_2 \}
= 
n_2^Q K^{n2}w^{c_2}\Prob\{ {\cal{P}}_2 \rightarrow {\cal{P}}_1 \}
\end{equation}
is then satisfied.
\end{description}

\item Case of vertex selection. We now suppose that we have selected a vertex,
which occurs with probability $(1-q)$. In this case, we attempt
either a double or a corner move, depending on the relative position
of the two bonds $(p_{i-1},p_i)$ and $(p_i,p_{i+1})$ shared by the chosen vertex.
For the relative orientations of figure \ref{f:doublecases}, 
a double move is attempted; for the
relative orientation of figure \ref{f:cornercases}, 
a corner move is attempted. In both cases the 
number of bonds remains constant ($\Delta n = 0$). For the transformations
$D_{III}$ and $C_{III}$ in which the number of crossings also remains unchanged 
(figure \ref{f:double2} and figure \ref{f:corner2}) the detailed balance holds 
clearly with
\begin{equation}
A(D_{III})=A(C_{III})=1
\end{equation}
by noticing that any such move has its own inverse by design. 
For the remaining cases (figure \ref{f:double1} and figure \ref{f:corner1}) 
the number of crossings
between ${\cal P}_1={\cal D}_1$ (or respectively ${\cal C}_1$) and
${\cal P}_2={\cal D}_2$ or $\overline{{\cal D}_2}$ (or respectively 
${\cal C}_2$ or $\overline{{\cal C}_2}$) varies by an amount $c_2-c_1 = 2$ 
and we simply have to account for the fact that there are two ways
to add the crossings while there is just one way to suppress them.
Setting $A(D_I)=A(C_I)=A^{(D,C)}_{+2}$ 
and $2A(D_{II})=A(C_{II})=A^{(D,C)}_{-2}$, we have
\begin{equation}
\Prob\{ {\cal{P}}_1 \rightarrow {\cal{P}}_2 \}  = 
A^{(D,C)}_{+2}
\times 
\left\{ \begin{array}{ll}
   2  \\
   1
\end{array} \right\}
\times
\frac{(1-q)}{n_1} \times
\left\{ \begin{array}{ll}
   \frac{1}{2}  \\
   1
\end{array}
\right\} \times \frac{1}{2}\times \min\left[1, w^2\right]
\end{equation}
where the first line is for a double move and the second line for a corner move.
The extra coefficients $2$ and $1/2$ in the case of a double move come 
respectively from the two different vertices which select the pair of parallel 
strands under consideration and from the probability $1/2$ to make the correct 
choice of direction $e_i$ in which we move the vertex $p_i$. 
We have conversely 
\begin{equation}
\Prob\{ {\cal{P}}_2 \rightarrow {\cal{P}}_1 \}  = 
\left\{ \begin{array}{ll}
\frac{A^{(D,C)}_{-2}}{2} \\ \scriptstyle{A^{(D,C)}_{-2}}
\end{array} \right\}
\times
\left\{ \begin{array}{ll}
2 \\ 1 \end{array} \right\}
\times
\frac{(1-q)}{n_2} \times \min\left[1, w^{-2}\right]
\end{equation}
The detailed balance
\begin{equation}
n_1^Q K^{n1}w^{c_1}\Prob\{ {\cal{P}}_1 \rightarrow {\cal{P}}_2 \}
= 
n_2^Q K^{n2}w^{c_2}\Prob\{ {\cal{P}}_2 \rightarrow {\cal{P}}_1 \}
\end{equation}
is then satisfied by choosing
\begin{equation}
A^{(D,C)}_{+2}=1;\quad A^{(D,C)}_{-2}=\frac{1}{2}.
\end{equation}
\end{enumerate}
\begin{table}
\small
\begin{tabular}{||r|r||r|r||r|r||r|r||}\hline
\multicolumn{2}{||c||}{Single Moves}&
\multicolumn{2}{c||}{Double Moves}&
\multicolumn{2}{c||}{Corner Moves}&
\multicolumn{2}{c||}{6-plaquette Moves}
 \\\hline
$S_I$            & $1$  & $D_I$     &$1$   & $C_I$    &$1$   & $R_I$ &$1$ \\
$S_{II}$         & $1$  & $D_{II}$  &$1/4$ & $C_{II}$ &$1/2$ & $$    &$$ \\
$S_{III}$        &$1/3$ & $D_{III}$ &$1$   & $C_{III}$&$1$   & $$    &$$ \\
$S_{IV},S_{IV}'$ &$1$   & $       $ &$ $   & $       $&$ $   & $$    &$$ \\
$S_{V},S_{V}'$   &$1/3$ & $       $ &$ $   & $       $&$ $   & $$    &$$ \\
\hline
\end{tabular}
\vspace{0.4cm}
\caption{Coefficients $A(M)$ for the different local moves.}
\label{t:lmoves}
\end{table}

The analysis above may seem a little cumbersome. In practice, the only
subtle point is the determination of the coefficients
$A(M)$, the list of which is displayed in Table \ref{t:lmoves}. 
These coefficients simply correct the fact that some 
transformations are in competition
(like a kink creation and a crossing creation) while the reverse transformation
has no competitor. As usual, the Metropolis criterion simply creates the 
correct ``energy" factor.  With the above choices for the acceptance 
coefficients 
$A(M)$, the transitions probabilities verify the detailed balance condition 
(\ref{detbal}) for all PSAP's ${\cal{P}}_1$, ${\cal{P}}_2$, with the invariant 
limit
distribution of eq.(\ref{distrorig})

\section{Improvements of the algorithm}

\subsection{Non-local Moves}
Algorithms based on local moves are known to have very long 
autocorrelation times which in turn lead to large statistical errors. 
For instance, it is known that, for the BFACF algorithm (which in our language
simply corresponds to forbid crossings), the so called ``exponential" 
autocorrelation time, which controls the relaxation of the Markov chain from an 
initial configuration to equilibrium, is infinite~\cite{ST89}. For
the same algorithm, the so called ``integrated" autocorrelation time, which controls 
the statistical error in measured quantities, is usually very large.

To remedy this problem, algorithms have been proposed~\cite{JW91a,CPS90} 
which combine local moves and non-local moves. The non-local moves
correspond to select two points far apart on the chain and to 
reverse the part of the chain between them. In these hybrid algorithms,
the non-local moves hopefully ensure the rapid equilibration within subspaces 
of fixed length $n$, while local moves ensure equilibration between different 
values of $n$. Now if one wants to study the statistics of polygons
{\sl with a fixed knot type}, one must make sure that {\sl the non-local moves 
preserve the topology of the polygon}. For the usual non-local moves performed
on 3-dimensional Self-Avoiding Polygons in $Z^3$, this is in general not
the case. One way to deal with this problem is then to check potential changes 
in the knot type by calculating a topological invariant such as the Alexander 
polynomial at every successful non-local move~\cite{JW91a}. 
This is however computationally costly and it also has the disadvantage that 
the Alexander polynomial is not a perfect characterization of the knot type
and distinct knots can have the same Alexander polynomial. 

\begin{figure}
\inseps{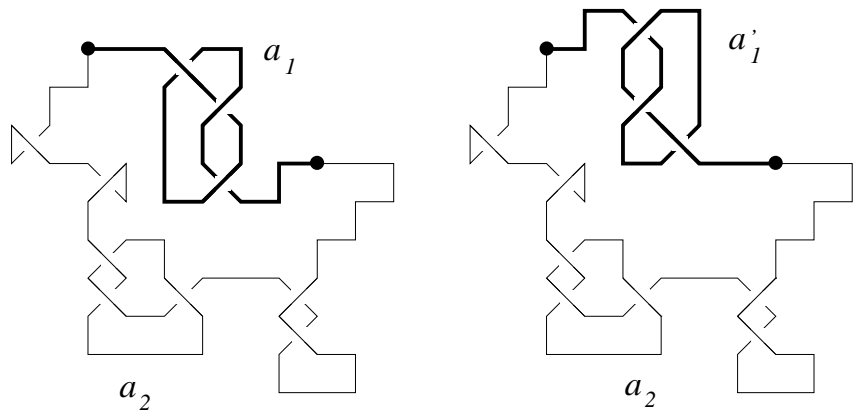}{0.8}
\caption{Non-local transformation of a PSAP. The arc $a_1'$ is the image
of $a_1$ by a rotation of $180^\circ$. The arc $a_2$ is unchanged. }
\label{f:nonlocal}
\end{figure}
\begin{figure}
\inseps{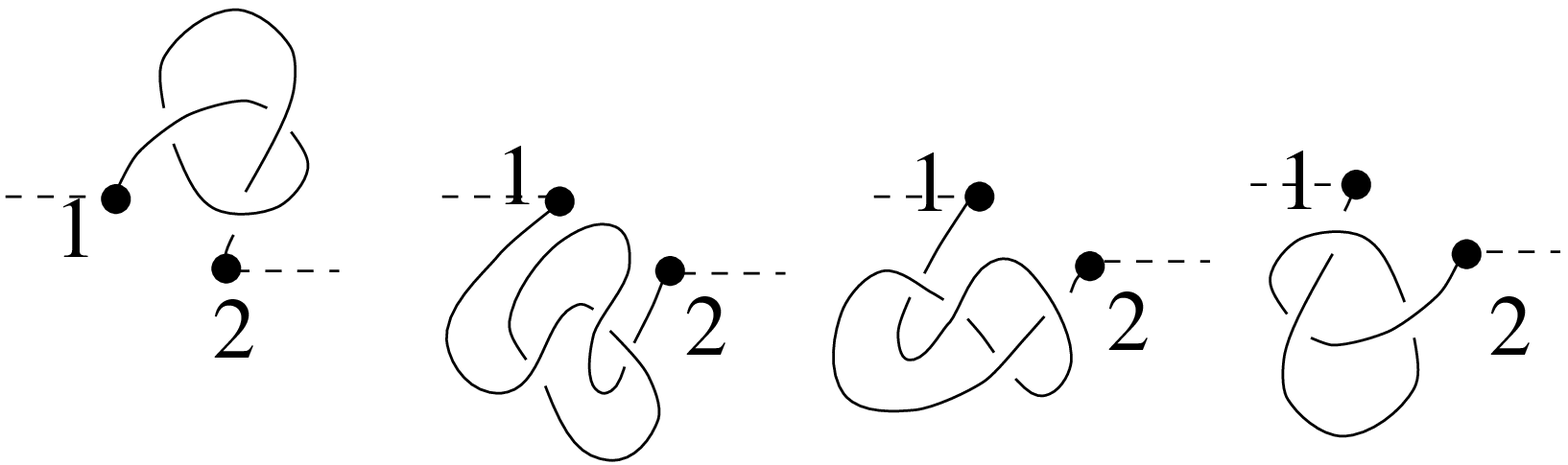}{0.8}
\caption{The passage from a trefoil to its image by a rotation of $180^\circ$
by use if local Reidemeister moves only. We indicate only the major
intermediate steps.}
\label{f:nonlocal3}
\end{figure}
\begin{figure}
\inseps{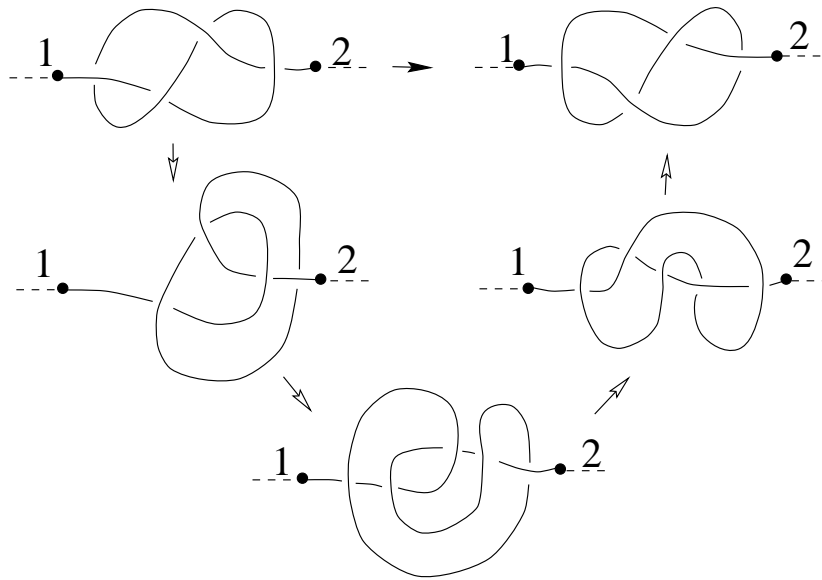}{0.8}
\caption{The passage from a figure eight to its image by a rotation of $180^\circ$
by use if local Reidemeister moves only. We indicate only the major 
intermediate steps.}
\label{f:nonlocal4}
\end{figure}

The situation is different in our model where the two-dimensional
character of the PSAP allows the introduction of non-local
moves that do preserve the topology of the PSAP.
The non-local move we consider are performed by first selecting
two vertices ($p_1$ and $p_2$) in the PSAP with uniform probability.
If these vertices are neighbors, no move is attempted; if not, the
vertices separate the PSAP into two arcs, say $a_1$ and $a_2$.
If the two arcs have no crossings in common we can perform a
rotation by $180^\circ$ of the shortest one, say $a_1$ (see figure 
\ref{f:nonlocal}). Let $a_1'$ be
the image of $a_1$ under this rotation. If $a_1'$ and $a_2$
do not intersect each other, no new crossing has been generated
and the move is accepted. The important feature of this move
is that it does not change the topology of the arc $a_1$ and therefore,
since we avoid topological interaction between the arcs, it preserves
the topology of the entire PSAP. To be more precise, the absence
of crossings between the two arcs guarantees that the knots in
$a_1$ and those in $a_2$ are well separated and the total knot can
be viewed as the compositions of the two knots. It is known that 
the composition of two knots can give at most two different results according 
to whether their (arbitrarily fixed) orientations do or do not match 
in the composition. Note that our transformation reverses this 
matching of orientations. The result is however unique as soon as at least 
one of the knots is reversible (i.e. its two opposite orientations
are in fact equivalent). This is the case for all the knots that we 
will use, which are made of trefoils ($3_1$) and figure eight ($4_1$) knots
only, both of which are reversible.
In practice, we can also argue that, for a trefoil ($3_1$) or figure eight ($4_1$)
knot, the passage from the arc $a_1$ to its image $a_1'$ can be done by using 
Reidemeister moves which {\sl do not involve} the arc $a_2$. This is illustrated 
for the trefoil knot in figure \ref{f:nonlocal3} and for the
figure eight knot in figure \ref{f:nonlocal4}. 

The non-local move described above is the simplest one that can be implemented. 
More sophisticated moves can be considered in which the constraint of having no 
crossings between $a_1$ and $a_2$ can be relaxed to weaker conditions.
We however limited ourself to the simplest non-local moves without crossings.

In our algorithm, we therefore have an additional free parameter $0\le p <1$
which is the probability that we attempt  non-local move rather than a local
move.

\subsection{Multiple Markov Chain Method}
 
Another way to improve the slow convergence of the local algorithm is by 
sampling along a set of several Markov chains run in parallel.

This method, refered to as the Multiple Markov
Chain (MMC) method, has been introduced by Geyer \cite{Geyer91} 
and then adapted to interacting self-avoiding walks with fixed 
length by Tesi et al~\cite{TJOW96} (see also~\cite{Orlandini96}  
for a more detailed analysis of the method).

We describe here a variant of this approach suitable for 
simulations in the grand canonical ($n$ varying) ensemble.
The starting point is the single Markov chain whose limiting distribution 
$\pi({\cal P})$ depends on the step and crossing fugacities $K$ and $w$.
Here we will consider a procedure with a fixed value of $w$, hence 
the $w$ dependence will be omitted in this section.
For values of $K$ close to (and below) the critical value $K_c$ where the 
average length of the PSAP diverges, the Markov chain samples configurations
with very large values of $n$ and the convergence of the algorithm is 
therefore very slow. On the contrary for a value $K'$ of the
step fugacity such that $K'<<K_c$, the sampled configurations have small 
values of $n$ and the convergence of the local algorithm is fast. 
The idea is to select a set of values $K'=K_1 < K_2 < K_3 < \ldots < K_m=
K $ to interpolate between $K'$ and $K$, so that $K_j$ and $K_{j+1}$
are close enough to ensure that there is a considerable overlap between
the distribution at $K_j$ and that at $K_{j+1}$.

The $m$ Markov chains at $K_1, K_2, \ldots, K_m$ are evolved
in parallel, and we let the chains interact by possibly exchanging
conformations as follows: select at random two neighboring chains with
fugacities $K_j$ and $K_{j+1}$.  A trial move is an attempt to swap 
the two current conformations of these chains. If we denote by
$\pi_j ({\cal P})$ the equilibrium weight of the state ${\cal P}$ for 
the chain at fugacity $K_j$, and ${\cal P}_j$ and ${\cal P}_{j+1}$ the 
current conformations in the $j$-th and $(j+1)$-th chain, we accept the 
trial move (i.e. swap ${\cal P}_j$ and ${\cal P}_{j+1}$) with the
acceptance rate
\begin{equation}
a({\cal P}_j \leftrightarrow {\cal P}_{j+1}) = 
\min \left( 1, \frac{\pi_j({\cal P}_{j+1}) \pi_{j+1}({\cal P}_j)}
{\pi_j({\cal P}_j) \pi_{j+1}({\cal P}_{j+1})}
\right) .
\label{MMC}
\end{equation}

Note that we do not need to compute the normalization of the limiting 
distributions (i.e. the partition functions) since these normalizations
drop out in the ratio above.  

The whole process is itself a Markov chain, which we can call a 
{\sl composite Markov chain}. Since the underlying Markov chains are 
ergodic (for a fixed knot type), so is the composite Markov chain. 
Moreover the composite chain is in detailed balance since the 
``swap"-move as well as the moves in the underlying chains are. 
Consequently, the invariant limiting distribution is the product
of the distributions of $m$ separate Markov chains at the temperatures 
$K_1 < K_2 < \ldots < K_m$.

One can clearly understand the advantage of this method by focusing on a
particular value $K_j$. Every successful swap involving this chain and the
chain $K_{j+1}$ corresponds to a big change in the configuration at
this value $K_j$, so the correlation time at each $K_j$ is markedly
reduced. Swaps move the system at a particular $K_j$ to new regions
of the configuration space and therefore tend to eliminate 
problems of quasi-ergodicity. If two chains $K_j$ and $K_{j+1}$ are far 
apart, the changes in the configuration are likely to be large but 
eq.(\ref{MMC}) suggests that bigger distances between the two selected 
chains lead to smaller acceptance rates for the swaps.
In practice, the set of the values of $K$ is chosen so that
successive values of $K$ are sufficiently close for swapping to occur 
rather frequently, and $K_1$ is small enough for convergence at $K_1$ 
to be rapid. All this requires that the total number $m$ of chains
is sufficiently large. Hence the gain in convergence rate is obtained at only
marginal cost in computer time.

\begin{table}
\small
\begin{tabular}{|r||r|r|r|r|r|r|r|r|}\hline
$K$ &
\multicolumn{2}{c||}{(1) local}&
\multicolumn{2}{c||}{(2) local + non-local}&
\multicolumn{2}{c||}{(3) local + MMC}&
\multicolumn{2}{c||}{(4) all}
\\
\cline{2-9}
 & $\langle n \rangle$ & $\tau (n)$
 & $\langle n \rangle$ & $\tau (n)$
 & $\langle n \rangle$ & $\tau (n)$
 & $\langle n \rangle$ & $\tau (n)$
 \\\hline
$0.3760$ &$76 \pm 4$ & $18 \pm 3$ &$75 \pm 3$ & $12 \pm 2$ 
          &$75 \pm 1$ & $2.2 \pm 0.2$ &$75 \pm 1$ & $1.5 \pm 0.2$ \\
$0.3771$ &$110 \pm 11$ & $50 \pm 10$ &$100 \pm 6$ & $20 \pm 5$ 
          &$111 \pm 4$ & $5.6 \pm 0.9$ &$112 \pm 2$ & $3.0 \pm 0.3$ \\
$0.3780$ &$160 \pm 19$ & $67 \pm 11$ &$188 \pm 18$ & $40 \pm 15$ 
          &$188 \pm 9$ & $14 \pm 3$ &$192 \pm 9$ & $13 \pm 3$ \\
$0.3783$ &$  $ & $   $ &$260 \pm 35$ & $90 \pm 25$
          &$231 \pm 17$ & $47 \pm 15$ &$259 \pm 19$ & $35 \pm 10$ \\
\hline
\end{tabular}
\vspace{0.4cm}
\caption{Integrated autocorrelation times for the average length 
$\langle n\rangle$ of the unknot  at $w=1$,  for (1) the purely local 
algorithm, (2) the local + non-local algorithm, (3) the local + 
MMC algorithm and (4) local + non-local + MMC algorithm.
Each run has a total of $10^8$ local moves and the unit 
of sampling is $10^4$ local moves. The data which are not reliable 
are omitted.
}
\label{t:unknot}
\end{table}

In Table \ref{t:unknot}  we compare our estimates of the
integrated autocorrelation time (given in units of sampling, i.e. per
$10^4$ attempted local moves in our tests) of the average PSAP lengths,
as a function of $K$, for the four different implementations of
the algorithm: (1) the purely local algorithm using only elementary
local moves, (2) the local + non-local algorithm, using both local
and non-local moves, (3) the local MMC algorithm, using local moves
only together with the MMC sampling and (4) the local + non local
MMC algorithm, using local and non local moves together with
the MMC sampling. In all cases, the integrated autocorrelation time 
for the polygon average length increases as $K$ approaches the critical 
value $K_c$, as expected. However, it is clear that both non-local
moves and MMC swaps lead to a significant decrease in this integrated 
autocorrelation time. 

\section{Results for $w=1$}

\begin{figure}[htbp]
\inseps{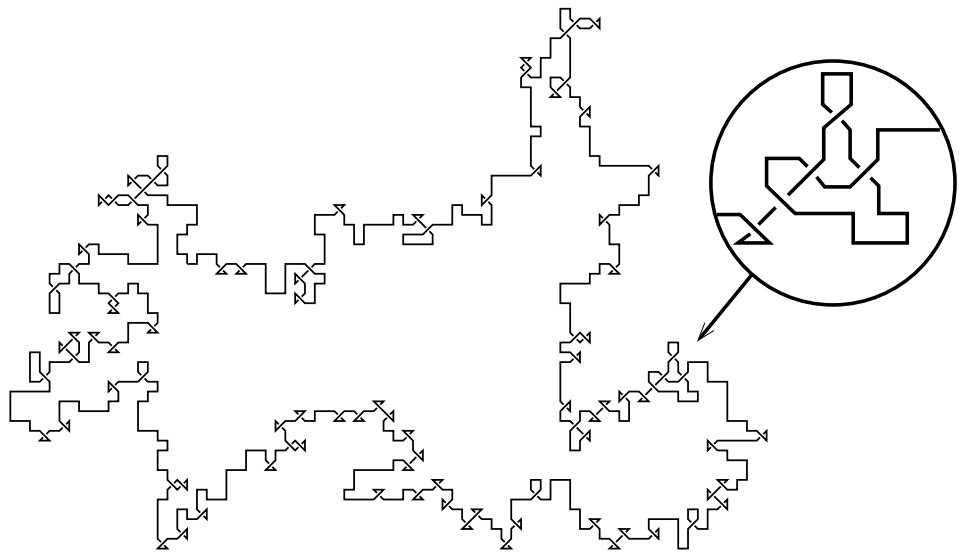}{0.7}
\caption{ A typical configuration for the $3_1$ at $w = 1$.
The number of bonds is $n=906$. We have indicated the location of
the knotted section.}
\label{f:31w1k7}
\end{figure}

In this section, we focus on the model with $w=1$, i.e. we do not
attach any particular weight to the crossings. 
A typical configuration with the topology of a trefoil knot is
shown in figure \ref{f:31w1k7}.

\subsection{Connectivity $\mu$ and Entropic Exponent $\alpha$}

We present here our estimates for the connectivity constant
$\mu$ of the PSAP and the entropic exponent $\alpha$ for $w=1$.
These quantities are defined by the large $n$ behavior of 
the {\sl number} $p_n(\tau)$ of configurations of fixed knot-topology 
$\tau$ with a fixed number $n$ of bonds, expected to be of the form
\begin{equation}
p_n(\tau)\sim A(\tau) \left(\mu(\tau)\right)^n \, n^{\alpha(\tau)-3}
\label{scalingform}
\end{equation}
In particular we study the 
dependence of $\mu(\tau)$ and $\alpha(\tau)$ on the topology 
$\tau$ of the PSAP. A method to compute $\mu(\tau)$ and $\alpha(\tau)$ 
relies on the asymptotic behavior for the the average length
$\langle n(\tau) \rangle $ of the PSAP for $K$ close to the critical 
fugacity $K_c(\tau)=(\mu(\tau))^{-1}$, namely:
\begin{equation}
\la n(\tau)\ra \approx \frac{[\alpha(\tau)+Q-2]K\mu(\tau)}{1-K\mu(\tau)}
\label{nscalgran}
\end{equation}
This behavior directly follows our choice of the statistical weight in
eq.(\ref{distrorig}), which yields for $w=1$
\begin{equation}
\la n(\tau)\ra = \frac{\sum_n n^{Q+1} p_n(\tau) K^n}{
\sum_n n^Q p_n(\tau) K^n}
\label{naver}
\end{equation}
Inserting eq.(\ref{scalingform}) directly leads to eq. (\ref{nscalgran}).
At leading order, we can use eq.(\ref{nscalgran}) to 
approximate $1/\langle n (\tau) \rangle$ as: 
\begin{equation}
\frac{1}{\la n(\tau)\ra} \approx \frac{1-K\mu(\tau)}{[\alpha(\tau)+Q-2]K
\mu(\tau)}
= \frac{1}{[\alpha(\tau) +Q-2]K\mu(\tau)} - \frac{1}{\alpha(\tau)+Q-2} .
\label{ninvscal}
\end{equation}
An estimate of $\mu(\tau)$ and $\alpha(\tau)$ can therefore 
be obtained by a linear extrapolation of $1/\langle n (\tau) \rangle$
as a function of $1/K$ for $1/\langle n(\tau) \rangle \to 0$.

\begin{figure}[htbp]
\inseps{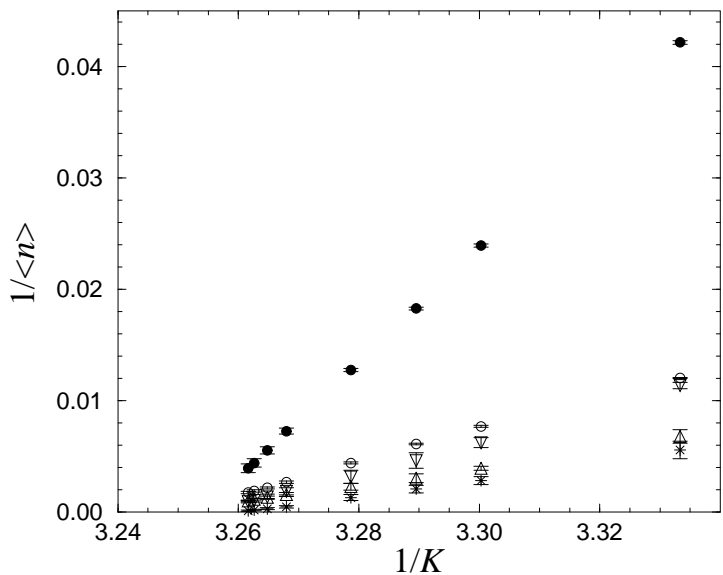}{0.6}
\caption{Plot of $1/\langle n \rangle$ as a function of $1/K$
for the unknot ($\bullet$), the trefoil ($\circ$), the figure eight ($\scriptstyle
\nabla$), the composite knot $3_1 \# 3_1$ ($\scriptstyle \Delta$) and the composite 
knot $3_1\# 4_1$ ($\scriptstyle{+\kern -6.75pt \times}$).}
\label{f:figure1}
\end{figure}

In figure~\ref{f:figure1} we have plotted $1/\la n(\tau)\ra$ as a 
function of $1/K$ for the topology of the unknot ($\emptyset$), that of the trefoil 
knot ($3_1$), that of the figure eight knot ($4_1$),
that of the composite knot made of two trefoils $(3_1 \# 3_1)$ 
and that of the composite knot made of a trefoil and a figure eight $(3_1 \# 4_1)$. 
These data are obtained with $Q=2$.
We note  that as the complexity of the topology increases, the corrections 
to the linear scaling (\ref{ninvscal}) become more and more important, 
but for $K$ sufficiently close to the critical value 
$K_c(\tau) = \mu(\tau)^{-1}$, a linear behavior is indeed obtained. 
By extrapolating the data, using the linear fit of  equation 
(\ref{ninvscal}), we obtain the following estimates: 
\begin{eqnarray}
\mu (\emptyset) &=&  3.254 \pm 0.01, \nonumber  \\
\mu (3_1)  &=& 3.250 \pm 0.06, \nonumber \\
\mu (4_1 ) &=& 3.255\pm 0.07, \\
\mu (3_1 \# 3_1) &=& 3.249\pm 0.13, \nonumber \\
\mu (3_1 \# 4_1) &=& 3.261\pm 0.15. \nonumber
\label{res.mu}
\end{eqnarray}
These values coincide up to the second decimal, and it seems reasonable
to assume that they are indeed all equal. In fact, since they were measured
from completely independent simulations, we can take their average 
to estimate the growth constant of PSAP of a fixed knot type: we obtain
$\mu(\tau)= \mu_1 = 3.251\pm 0.002$ ($95$\% confidence interval)
for PSAP's with arbitrary fixed topology $\tau$. 
From the same linear fit an estimate of the entropic exponent $\alpha$
can also be obtained:
\begin{eqnarray}
\alpha (\emptyset) &=&  0.58 \pm 0.07, \nonumber  \\
\alpha (3_1)  &=& 2.10 \pm 0.08, \nonumber \\
\alpha (4_1 ) &=& 2.13\pm 0.10, \\
\alpha (3_1 \# 3_1) &=& 3.91\pm 0.24, \nonumber \\
\alpha (3_1 \# 4_1) &=& 4.05\pm 0.25. \nonumber
\label{res.alpha}
\end{eqnarray}
In contrast with $\mu$, the exponent $\alpha$ depends clearly on the topology. 
However, this dependence appears to be weak in the sense that $\alpha$ depends 
only on how many primary knots form the knot under consideration
and not on the precise nature of these knots, hence $\alpha (3_1)=\alpha (4_1 )$
and $\alpha (3_1 \# 3_1)=\alpha (3_1 \# 4_1)$ within numerical uncertainties.
Such a result was already found in \cite{OTJW96,OTJW98} for three-dimensional
knotted polygons. There however, it was found that $\alpha(\tau)$ increases by 
one for each added primary knot, leading to $\alpha(\tau)=\alpha(\emptyset)
+N_f$ where $\tau$ is a knot with $N_f$ prime factors. This formula
does not seem to hold in our case. 

Another way to estimate $\mu$ and $\alpha$ relies on the
maximum likelihood method as suggested in ~\cite{BS85}. In this case,
we assume that for $n>n_{min}$ the relation
\begin{equation}
p_n = \mu^n n^{\alpha -3} \left (1 + a/n\right )
\end{equation}
holds, where $a$ is a free parameter that we can vary
to mimic a correction to scaling. By varying $n_{min}$ and $a$
we can probe the sensitivity of our estimates to scaling.
We generally find a region of minimum sensitivity from which we
extract the following estimates:
\begin{eqnarray}
\mu (\emptyset) &=&  3.254 \pm 0.005, \nonumber  \\
\mu (3_1)  &=& 3.251 \pm 0.07, \nonumber \\
\mu (4_1)  &=& 3.253 \pm 0.09, \\
\mu (3_1 \# 3_1) &=& 3.245\pm 0.10,\nonumber \\
\mu (3_1 \# 4_1) &=& 3.255 \pm 0.11. \nonumber
\label{rescan.mu}
\end{eqnarray}
and for the entropic exponents:
\begin{eqnarray}
\alpha (\emptyset) &=&  0.58 \pm 0.03, \nonumber  \\
\alpha (3_1)  &=& 1.86 \pm 0.06, \nonumber \\
\alpha (4_1)  &=& 1.95 \pm 0.08,  \\
\alpha (3_1 \# 3_1) &=& 4.20\pm 0.15, \nonumber \\
\alpha (3_1 \# 4_1) &=& 3.90\pm 0.20. \nonumber
\label{rescan.alpha}
\end{eqnarray}
All these results are in agreement with the estimates (\ref{res.mu}) and
(\ref{res.alpha}).
\subsection{The Metric Exponent $\nu$}

For the mean square radius of gyration, we assume the
large $n$ behavior
\begin{equation}
\langle R^2 (\tau)\rangle_n = A_{\nu}(\tau)n^{2\nu(\tau)}
\label{r2scalknot}
\end{equation}
where the average is performed over configurations with fixed $n$.
In order to estimate $\nu$ from our data in the grand-canonical ensemble, 
we compute the expected value of the mean square radius of gyration, 
\begin{equation}
\la R^2(\tau)\ra = \frac{\sum_n \langle R^2(\tau)\rangle_n n^Q p_n(\tau) 
K^n}{\sum_n n^Q p_n(\tau) K^n}
\label{expr.R2}
\end{equation}
Using the asymptotic forms (\ref{scalingform}) and (\ref{r2scalknot}), we 
get the following approximation for $\la R^2(\tau)\ra $ 
close to the critical $K_c$: 
\begin{equation}
\la R^2(\tau) \ra \simeq h(\tau)\left(1-K\mu(\tau)\right)^{-2\nu(\tau)} 
\label{r2scalgran1}
\end{equation}
or equivalently, using eq.(\ref{nscalgran}), the approximation:
\begin{equation}
\la R^2(\tau) \ra \simeq h'(\tau)\langle n (\tau)\rangle^{2\nu(\tau)} 
\label{r2scalgran2}
\end{equation}
where we now have a relation between $\la R^2(\tau)\ra$ and $\la n(\tau) \ra$.  
A $\log$-$\log$ fit of our data to equation (\ref{r2scalgran2}), 
where $K$ takes a different value for each chain in our multiple Markov chain 
Monte Carlo, can be used to estimate $\nu$ for each knot type.  Our best 
estimates are
\begin{eqnarray}
\nu(\emptyset) = 0.747 \pm 0.005 , \nonumber \\
\nu(3_1) = 0.758\pm 0.010 , \nonumber \\
\nu(4_1) = 0.748\pm 0.010 ,  \\
\nu(3_1\#3_1) = 0.760\pm 0.020, \nonumber \\
\nu(3_1\#4_1) = 0.720 \pm 0.040. \nonumber 
\label{nuestgc1}
\end{eqnarray}
Within the stated error bars, the results above are all identical
and consistent with the {\sl exact} value $\nu_{\rm SAW}=3/4$ \cite{Nien82}
for self-avoiding walks or self-avoiding polygons in $d=2$. 
This seems to indicate that although polygons in our model 
are not strictly self-avoiding since crossings are allowed for
$w=1$, the presence of these crossings limited to diagonal edges
does not modify the scaling behavior of the radius of gyration. 

\section{Results for $w << 1$}

It is interesting to consider our model in the limit $w\to 0$ 
where crossings become forbidden. At $w=0$, our model coincides
with that of a strictly self-avoiding polygon on the square lattice.
Still, we can start from a configuration with a non-trivial knot
topology $\tau$ (say that of the trefoil knot) with its minimal number
$c_{\rm min}(\tau)$ of crossings ($c_{\rm min}(3_1)=3$ for the trefoil) 
and look at the limit $w\to 0$ of $G_\tau(Q,K,w)/w^{c_{\rm min}(\tau)}$. 
In this case, we explore the phase space of all the configuration
of knot type $\tau$ and with the minimal number of crossings 
$c_{\rm min}(\tau)$. This limit is the ``minimal" extension of the
strictly self-avoiding model to include a non-trivial topology.
In this limit, we of course expect the model to be in the universality 
class of strictly self-avoiding walks. 

\subsection{Results for $\mu$ and $\alpha$}

In practice, the easiest way to study the $w\to 0$ limit is simply
to fix in our simulation a very small value of $w$. The results 
presented here have been obtained with the value of the crossing
fugacity $w$ fixed to $w=0.0001$.
Here again we define the connectivity constant $\mu$ and the
entropic exponent $\alpha$ for the asymptotic behavior of the {\sl number} 
$p_n^{(0)}(\tau)$ of configurations of polygons in the topology
$\tau$, with $n$ bonds and with the minimal number $c_{\rm min}(\tau)$
of crossings compatible with $\tau$. As for the case $w=1$, we can 
estimate  $\mu$ and $\alpha$ by plotting $1/\la n(\tau)\ra$ as a 
function  of $1/K$ and by making a linear extrapolation for 
$1/\la n(\tau)\ra \to 0$.

For $\mu$ we obtain the following estimates:
\begin{eqnarray}
\mu (\emptyset) &=&  2.640 \pm 0.01, \nonumber  \\
\mu (3_1)  &=& 2.640 \pm 0.01, \nonumber \\
\mu (4_1)  &=& 2.639 \pm 0.02,  \\
\mu (3_1 \# 3_1) &=& 2.641\pm 0.05, \nonumber \\
\mu (3_1 \# 4_1) &=& 2.645 \pm 0.08 . \nonumber
\label{res.mu0}
\end{eqnarray}
These values coincide to the second decimal place, and it seems reasonable
to assume that they are indeed all equal. It is interesting to notice that
these estimates of $\mu(\tau)$ are a good upper bound for
the best available estimates of the connectivity constant
for self-avoiding polygons in the $d=2$ square lattice ($\mu_0 = 2.63815$)
\cite{CPS90b}.

For the entropic exponents we obtain:
\begin{eqnarray}
\alpha (\emptyset) &=&  0.503 \pm 0.040, \nonumber  \\
\alpha (3_1)  &=& 2.24 \pm 0.10, \nonumber \\
\alpha (4_1)  &=& 2.28 \pm 0.10,  \\
\alpha (3_1 \# 3_1) &=& 3.2\pm 0.3, \nonumber \\
\alpha (3_1 \# 4_1) &=& 3.3 \pm 0.5 . \nonumber
\label{res.alpha0}
\end{eqnarray}

As for the case $w=1$, we again see a dependence of $\alpha$ 
on the number of prime factors forming the knot. We note
however that the values of $\alpha$ for a fixed knot type 
are different in this limit ($w\to 0$) from the $w=1$ values.
For the unknot, we recover the exact value $\alpha_{\rm SAW}=1/2$ 
\cite{Nien82}. This is expected since strictly self-avoiding rings in two 
dimensions cannot form knots, thus the usual $\alpha$ exponent derived
form field theory corresponds to fixing the topology to that of the unknot
rather than summing over all topologies as in three dimensions. 

\subsection{The Metric Exponent $\nu$}

As for the case $w=1$, we have evaluated the exponent $\nu$ in
the limit $w\to 0$ from the relation of eq. \ref{r2scalgran2}
between the average radius of gyration $\langle R^2(\tau)\rangle$ 
and the average number of bonds $\langle n (\tau) \rangle$. Our
estimates are 
\begin{eqnarray}
\nu(\emptyset) = 0.748 \pm 0.015 , \nonumber \\
\nu(3_1) = 0.753\pm 0.010 , \nonumber \\
\nu(4_1) = 0.749\pm 0.011 , \\
\nu(3_1\#3_1) = 0.736\pm 0.015, \nonumber \\
\nu(3_1\#4_1) = 0.742 \pm 0.020. \nonumber 
\label{nuestgc2}
\end{eqnarray}

We thus obtain an exponent $\nu$ independent of the knot-type
and in perfect agreement with the exact value $\nu_{\rm SAW}=3/4$ for 
self-avoiding walks in two dimensions.

\subsection{Statistics of Arc Lengths: Localization of Knots.}

One advantage of taking this limit $w\to 0$ is that, by preventing
the creation of spurious crossings, we have a fixed minimal number
$c_{\rm min}(\tau)$ of crossings which we can easily localize along 
the PSAP. In particular, we can study the typical ``size" of the knot 
in a given topology and see for instance if primary knots tend
to be localized on a small number of bonds or prefer to extend over
the whole PSAP. A simple way to analyze this phenomenon is to study the
distribution on the lengths of the arcs between consecutive crossings.

In the case of a trefoil knot, we have exactly $c_{\rm min}(3_1)=3$
crossings which separate exactly $6$ arcs along the PSAP. For each
configuration with $n$ bonds, we can classify these arcs in order of 
decreasing arc length $l_i=\lambda_i n$ with 
$\lambda_1\ge\lambda_2\ge\cdots\ge\lambda_6$ and $\sum_{i=1}^6 \lambda_i=1$. 
Here $\lambda_i$ is thus the fraction of the total length $n$ inside
the $i$-th largest arc.

\begin{figure}[htbp]
\inseps{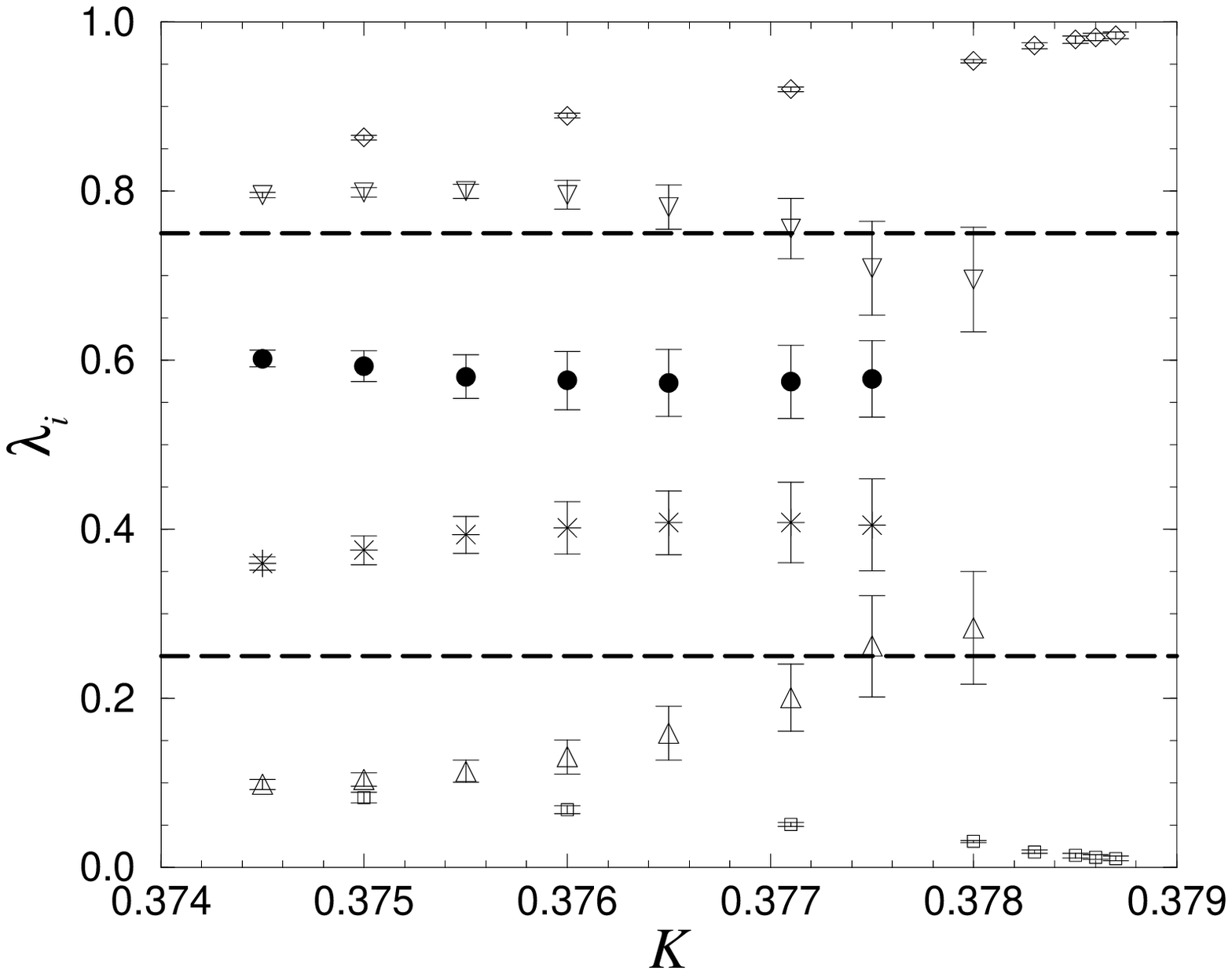}{0.6}
\caption{Average length fraction $\lambda_1$ and $\lambda_2$ of the two 
largest arcs along the PSAP as a function of $K$ for the trefoil 
($\lambda_1 \to \diamond$,  $\lambda_2 \to {\scriptstyle \Box}$),
the composite knot $3_1 \# 3_1$ ($\scriptstyle \nabla$ and $\scriptstyle 
\Delta$) and the composite knot $3_1 \# 4_1$ ($\bullet$ and 
$\scriptstyle{+\kern -6.75pt \times}$).
The dashed lines indicate the values $3/4$ and $1/4$ expected for
two strictly localized knots moving freely along the PSAP.}
\label{f:figcro}
\end{figure}

\begin{figure}[htbp]
\inseps{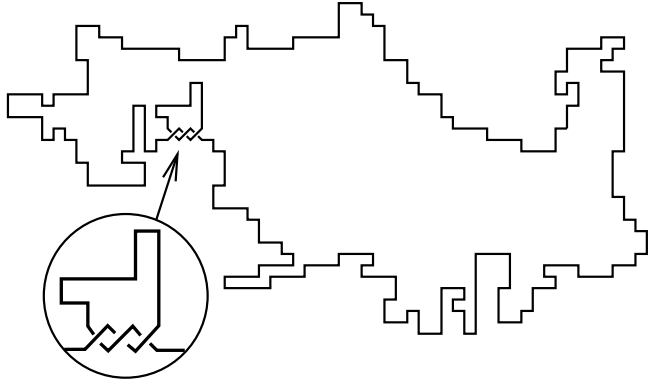}{0.55}
\caption{ A typical configuration for the trefoil knot ($3_1$) at $w\sim 0$.
The number of bonds is $n=312$. We have indicated the location of the
knotted section.}
\label{f:31w0k2}
\end{figure}

In figure ~\ref{f:figcro}, we have plotted the fraction of length
occupied by the two largest arcs for a trefoil knot, as a function 
of $K$. We see that as $K\to K_c$ ($\la n \ra \to \infty$), 
$\lambda_1\to 1$ while $\lambda_2$, and thus all the other $\lambda_i$
for $i>2$, tend to zero. This result is consistent with a {\sl localized}
knot which takes place over a number a bonds which is either
finite, or at least growth less rapidly than $n$. This result of localization
is corroborated by a direct visualization of a typical configuration, as shown 
in figure \ref{f:31w0k2}.

\begin{figure}[htbp]
\inseps{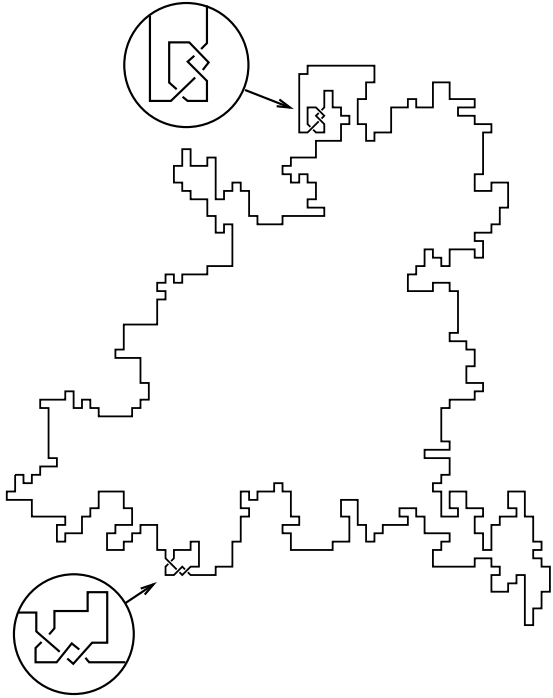}{0.5}
\caption{ A typical configuration for the composite knot $3_1\# 3_1$ at $w\sim 0$.
We have indicated the location of the two primary knots. }
\label{f:3131w0k1}
\end{figure}

Another interesting case is that of composite knots $3_1\#3_1$
and $3_1 \# 4_1$, which have respectively $c_{\rm min}(3_1\#3_1)=6$ and 
$c_{\rm min}(3_1\#4_1)=7$. These knots are thus made respectively of exactly 
$12$ and $14$ arcs. We have plotted on figure \ref{f:figcro} the proportions
of the total length $n$ occupied by the two largest arcs in both cases.
Although the statistical errors are rather important when $K\to K_c$, we see 
in each case that these proportions add up to one, with the proportion 
occupied by all the smaller arcs tending to zero. This result is consistent 
with a picture of the composite knot as made of two perfectly 
localized primary factors. The localization of the two primary knots is 
also directly visible on a typical configuration of the PSAP 
(see figure \ref{f:3131w0k1}).  If we assume that these localized factors
move freely along the PSAP and interact only locally by preventing to pass 
through one another (which would require creating additional crossings),  
we should expect that the proportions tend to $3/4$ and $1/4$. Indeed, for 
two points at random drawn on a loop of size $n$, the largest arc between 
them has average length $(1/n)\int_0^n dx \max(x,n-x)=(3n/4)$ and the 
complementary arc has thus average length $n/4$. 
In practice, we observe a value a $\lambda_1<3/4$ and $\lambda_2>1/4$,
which tends to indicate that the primary knots repel each other.
This repulsion might however be a finite size effect. Indeed the deviation
from the expected asymptotic values $3/4$ and $1/4$ is larger 
for the $3_1 \# 4_1$ knot than for the $3_1\# 3_1$ knot as the $4_1$ prime 
factor typically requires more bonds to be formed. 

\section{Results for $w >1$}

Up to now, we have studied only the case $w=1$ where no fugacity
is associated to the number of crossings and the case $w\to 0$ where
we have a minimal number of crossings compatible with the knot topology 
at hand. Although we obtained different values for the (non universal)
connectivity constant $\mu$, we found in both cases a value of 
$\nu$ compatible with the exact result $\nu_{\rm SAW}=3/4$ in two dimensions, 
which suggest that these two cases are in the universality class of
self-avoiding walks, and so is probably any case of the model with a 
value of $w$ between $0$ and $1$. In this section, we will rather
be interested in the behavior of the model for $w>1$.

\subsection{Connectivity $\mu$}
 
Let us first concentrate on the connectivity constant $\mu$.
If we denote by $p_{n,c}(\tau)$ the number of configurations
of PSAP with $n$ bonds, $c$ crossings and the knot topology $\tau$,
we can introduce the {\sl canonical partition function}
\begin{equation}
p_n(w,\tau)=\sum_{c\ge c_{\rm min(\tau)}}p_{n,c}(\tau)w^{c-c_{\rm min}(\tau)}
\end{equation} 
For $w=1$, we recover the {\sl number} $p_n(\tau)$ of configurations
with knot-topology $\tau$ and with $n$ bonds, irrespectively of their 
number of crossings. For $w=0$, we recover the {\sl number} 
$p_n^{(0)}(\tau)$ of configurations with topology $\tau$, number
of bonds $n$, and with the minimal number of crossings 
$c_{\rm min}(\tau)$.

We define $\mu(w,\tau)$ and $\alpha(w,\tau)$ by the large $n$ behavior
\begin{equation}
p_n(w,\tau)\sim A(w,\tau)\mu(w,\tau)^n n^{\alpha(w,\tau)-3}
\end{equation}

In the grand canonical ensemble, we have
\begin{equation}
G_\tau(Q,K,w)/w^{c_{\rm min}(\tau)}=\sum_{n}p_n(w,\tau)K^n n^Q
\end{equation}
hence we can again relate $\mu(w,\tau)$ to the inverse of the
critical value $K_c(w)$ of $K$ for which $\la n (w,\tau)\ra $
diverges.

Before we present our numerical results, let us first 
establish a lower and an upper bound for $\mu(w)$.

\begin{figure}
\inseps{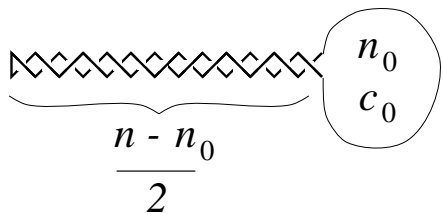}{0.5}
\caption{The growing of a linear sequence of crossings from
a compact knot.}
\label{f:growing}
\end{figure}

\noindent{\bf Lower bound for $w>1$:}
The canonical partition function $p_n(w,\tau)$ is clearly bounded
from below by the contribution $w^c$ of any acceptable configuration
in the given knot-topology class. Since $w>1$, we can obtain
a larger lower bound by selecting a configuration with the maximum
number of crossings. To obtain such a configuration, we start from
any particular realization of the knot, as compact as possible.
It will have, say, $n_0$ bonds and $c_0$ crossings. On any outermost
bond, we can start to grow a linear sequence of crossings, such 
as the one depicted in figure \ref{f:growing}. Note that we can
arbitrarily choose any of the two possible crossings for each
new crossing added in linear piece. The addition of this
linear sequence clearly does not modify the topology of the knot. 
If $n$ is the total number of bonds in the configuration, we get
a total number of crossings equal to $c_0+(n-n_0)/2$ for any of the
$2^{(n-n_0)/2}$ such configurations.
We thus get
\begin{equation}
p_n(w,\tau)> 2^{-\frac{n_0}{2}}w^{c_0-\frac{n_0}{2}-c_{\rm min}(\tau)}
\times (2w)^{\frac{n}{2}}
\end{equation}
In the large $n$ limit, this directly leads to the lower bound
\begin{equation}
\mu(w)\ge \sqrt{2w}
\label{lowerbound}
\end{equation}

\vskip .5cm 
\noindent{\bf Upper bound for $w>1$:}
It is clear that any configuration with $n$ bonds can have at most
$n/2$ crossings, irrespectively of its knot type.
We thus get for $w>1$
\begin{equation}
p_n(w,\tau)< p_n(\tau) w^{\frac{n}{2}}
\end{equation}
This directly leads to
\begin{equation}
\mu(w,\tau)\le \mu_1\sqrt{w}
\label{upperbound}
\end{equation}

From eqs.(\ref{lowerbound}) and (\ref{upperbound}), we can deduce that
at large $w$, $\mu(w)$ behaves like
\begin{equation}
\mu(w,\tau)\buildrel {w\to \infty} \over \sim \lambda \, \sqrt{w}
\label{asymmu}
\end{equation}
with $\sqrt{2}\le \lambda \le \mu_1\sim 3.251$.

\vskip .5cm
To complete our analysis, we can also give lower and upper bounds
on the case $0<w<1$. In this case, it is clear that
\begin{equation}
p_n^{(0)}(\tau)<p_n(w,\tau)< p_n(\tau) 
\end{equation}
and therefore
\begin{equation}
2.638\sim \mu_0\le\mu(w,\tau)\le \mu_1 \sim 3.251
\end{equation}

\begin{figure}[htbp]
\inseps{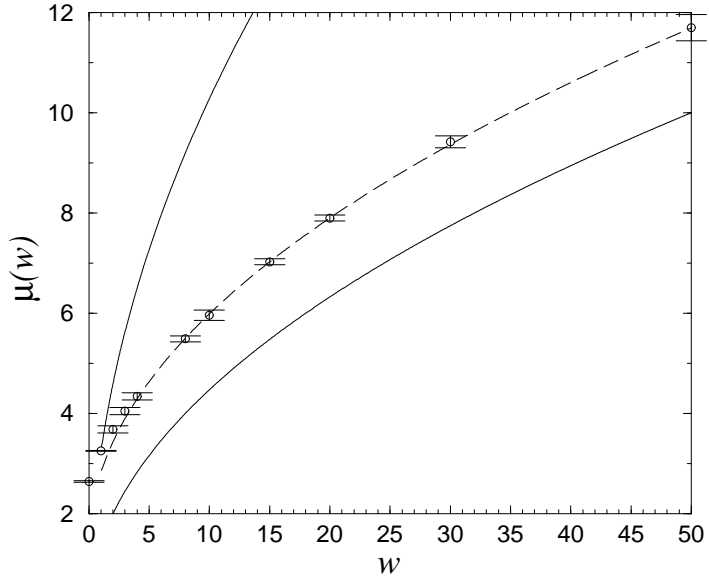}{0.6}
\caption{Estimate of $\mu$ as a function of the crossing fugacity $w$;
      the solid lines correspond respectively to the curves
      $\mu = 3.251 {w}^{\frac{1}{2}}$ (upper bound) and 
      $\mu = (2w)^{\frac{1}{2}}$ (lower bound). The dashed line corresponds
      to the fit $\mu=1.46(w^{\frac{1}{2}}-1)+2.86$.
}
\label{f:figmu}
\end{figure}

Figure \ref{f:figmu} presents our results for $\mu(w)$ together
with the above lower and upper bounds. At large $w$, we confirm 
the behavior of eq.(\ref{asymmu}) with $\lambda\simeq 1.46$ close 
to the lower bound $\sqrt{2}$. 

\subsection{The Metric Exponent $\nu$: Branched Polymer Behavior}

\begin{figure}[htbp]
\inseps{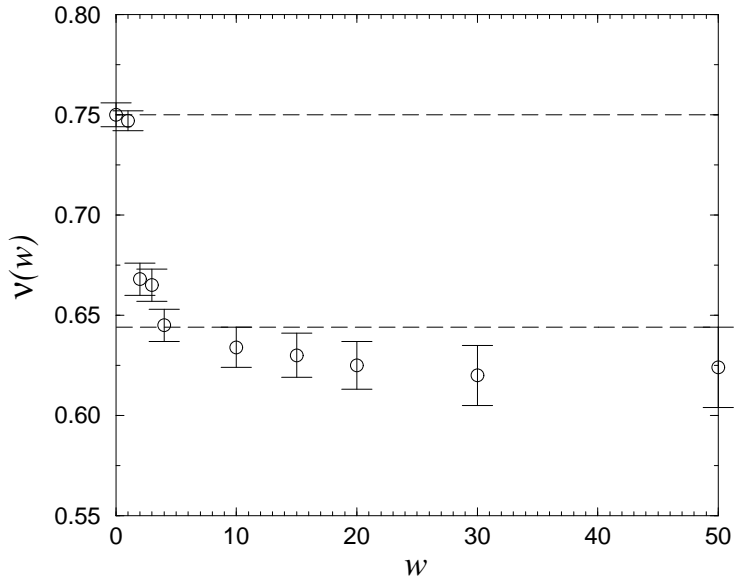}{0.6}
\caption{The exponent $\nu$ as a function of $w$. The dashed lines indicate
the values of $\nu$ for self-avoiding walks and branched polymers.}
\label{f:nuvsw}
\end{figure}

\begin{figure}[htbp]
\inseps{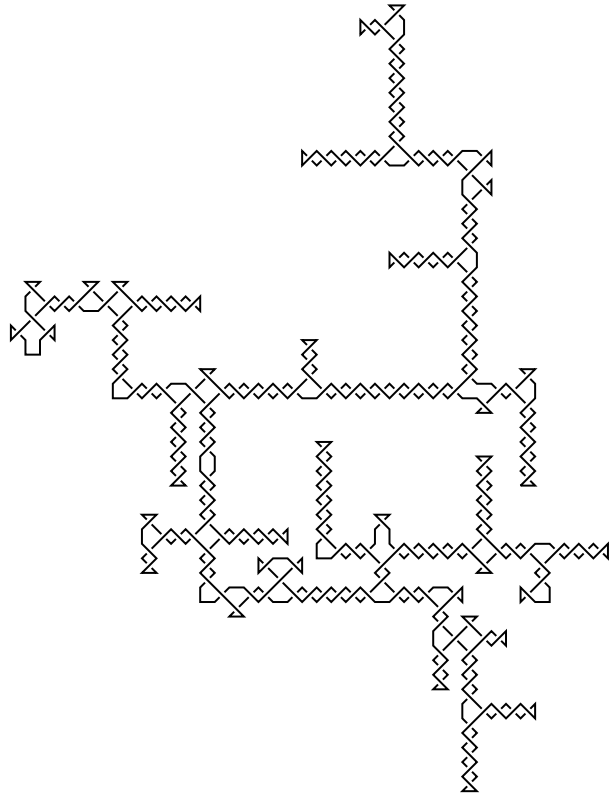}{0.5}
\caption{ A typical configuration for the unknot ($\emptyset$) at $w = 30$.
}
\label{f:figw30k7}
\end{figure}

\begin{figure}[htbp]
\inseps{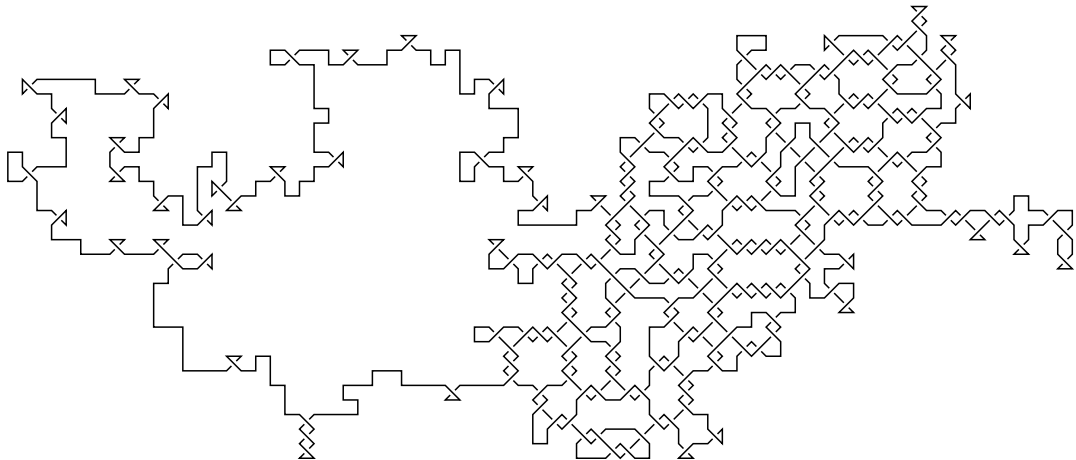}{0.8}
\caption{A configuration for the unknot at $w = 1$, with $n=1250$ bonds. 
One clearly sees two domains with very different densities of crossings, 
one being sparse while the other is dense.}
\label{f:unw1k5}
\end{figure}

We have also estimated the metric exponent $\nu$ for different values
of $w>1$.
Our best estimates are
\begin{eqnarray}
\nu(w=2) = 0.668 \pm 0.008 , \nonumber \\
\nu(w=3) = 0.665\pm 0.007 , \nonumber \\
\nu(w=4) = 0.645\pm 0.008 , \nonumber \\
\nu(w=10) = 0.634\pm 0.010 ,  \\
\nu(w=15) = 0.630\pm 0.011 , \nonumber \\
\nu(w=20) = 0.625\pm 0.012 , \nonumber \\
\nu(w=30) = 0.620\pm 0.015 , \nonumber \\
\nu(w=50) = 0.624\pm 0.020 , \nonumber
\label{nuestwgt1}
\end{eqnarray}
These values are plotted in figure \ref{f:nuvsw}, together 
with the values estimated for $w\to 0$ and $w=1$. Within the
error bars, we distinguish two values for $\nu$. For $w\le 1$, our
data are consistent with the value $\nu_{\rm SAW}=3/4$ of self-avoiding
walks. For $w>1$ however, our estimates are far below this value
and approximately located around the value $\nu_{\rm BP}\sim 0.64$
obtained for branched polymers in $D=2$ \cite{DDS82,JM92}. 
At large $w$, the branched polymer picture is corroborated by looking at a
typical configuration, as the one depicted in figure \ref{f:figw30k7}.
Such configurations are naturally selected at large $w$ since 
long one-dimensional coils maximize the number of crossings and
are thus energetically favored, while branching points, occuring at 
a finite energy cost, are favored for entropic reasons.
From our data, a transition from the self-avoiding behavior to
a branched polymer type statistics seems to occur at $w=1$. 
For $w<1$, crossings are penalized and play a marginal role
without changing the asymptotic behavior of the polymer, which
can be considered as self-avoiding. On the contrary, for $w>1$,
a finite density of crossings is present in a typical configuration,
which degenerates into a branched polymer conformation. At the 
transition  point $w=1$, we find occasionally  interesting configurations
showing a phase separation between a phase rich in crossings and
a phase with few crossings (see figure \ref{f:unw1k5}). 
In this case, the largest of the largest value $\nu=\nu_{\rm SAW}$ 
is observed. Finally, let us note that a similar transition from 
a self-avoiding behavior to a branched polymer behavior has already
been observed in an interacting two-tolerant trail model on the square 
lattice \cite{OSST92}.
 
\section{Conclusions}
In this paper, we introduced a two-dimensional lattice model 
of Projected Self-Avoiding Polygons, describing two-dimensional
knotted polymers rings with a fixed knot topology. We made a 
numerical study of the model using a Monte Carlo algorithm based 
on both local and non-local moves which preserve the knot topology 
of the polygon, together with a Multiple Markov Chain procedure. 
In the absence of a rigorous proof of the ergodicity of our
algorithm within a fixed knot type, we gave strong arguments in favor
of this ergodicity, together with the guidelines of a 
possible proof.
We presented our numerical results for the connectivity constant $\mu$ 
and the critical exponents $\nu$ and $\alpha$ for various topologies and 
various values of the fugacity $w$ associated to the number
of crossings in the projected polygon. The connectivity $\mu$ is found to 
depend only on this fugacity $w$ and not on the particular knot topology 
at hand. This is somewhat expected since $\mu$ typically measures
a local property which is not affected by a global constraint such
as fixing the knot topology. By measuring the radius of gyration of
the polygon, we observe only two possible values for the exponent $\nu$: 
a self-avoiding value $3/4$ found at low $w\le 1$ irrespectively of
the knot type, and a large branched polymer value $\simeq 0.64$ found
at large $w>1$ for all knot topologies. Our data suggest that the
transition between the two regimes occurs precisely at $w=1$. 
Finally, at fixed $w$, the exponent $\alpha$ for the number of accessible
configurations is found to depend only on the number of prime factors 
forming the knot. A weak $w$ dependence is also observed.

One nice feature of our two-dimensional model is the possibility
to have more specific characterizations of the polygon conformations,
such as the number of crossings or the arc lengths between successive 
crossings along the chain. This allowed us in particular to measure
the typical extension of a knot inside the polygon. In the limit 
$w\to 0$, we found for instance that the prime factors in a knotted 
polygon are localized and weakly interacting. 

Other applications of this model can be imagined, such as 
a statistics of the knot type of randomly drawn polygons.
For an arbitrary polygon, the determination of the knot topology 
could be simply achieved in principle by imposing a very low value 
of $w$ and letting the system evolve toward a configuration with
a minimal number of crossings in the projection.
In particular, it would be interesting to understand how the 
knot complexity increases with the number of monomers in the chain.

\end{document}